\documentclass{iopart}
\usepackage{iopams}
\usepackage{graphicx}
\usepackage[square,numbers,sort&compress]{natbib}

\def\aap{{\it Astron.~\& Astrophys.}}
\def\aj{{\it Astron.~J.}}
\def\apj{{\it Astrophys.~J.}}
\def\apjl{{\it Astrophys.~J.~Lett.}}
\def\apjs{{\it Astrophys.~J.~Supp.}}
\def\araa{{\it Ann.~Rev.~Astron.~\& Astrophys.}}
\def\nat{{\it Nature}}
\def\mnras{{\it Mon.~Not.~Roy.~Astron.~Soc.}}
\def\physrep{{\it Phys.~Rep.}}
\def\prd{{\it Phys.~Rev.~{\rm D}}}
\def\phz{{photo-{\it z}}}
\def\Phz{{Photo-{\it z}}}

\def\mpc{\mbox{$h^{-1}$Mpc}}

\newcommand{\be}{\begin{equation}}
\newcommand{\ee}{\end{equation}}
\newcommand{\bea}{\begin{eqnarray}}
\newcommand{\eea}{\end{eqnarray}}

\begin{document}

\title[Cosmic tomographies]{Cosmic tomographies: baryon acoustic
oscillations and weak lensing}

\author{Hu Zhan}
\address{Department of Physics, University of California, 
Davis, CA 95616, USA}
\ead{zhan@physics.ucdavis.edu}

\begin{abstract}
We explore the complementarity between two tomographic probes of the
universe: baryon acoustic oscillations (in the galaxy power spectrum) 
and weak gravitational lensing. The galaxy power spectrum 
characterises the density fluctuations, whereas the weak lensing 
shear power spectrum is a direct measure of
the potential fluctuations. We find that photometric measurements of
baryon oscillations alone do not provide very tight constraints on 
the dark energy equation of state parameters, partially due to our 
uncertain knowledge of the galaxy clustering bias. Weak lensing, on 
the other hand, is adversely impacted by the uncertainties of the 
probability distribution of photometric redshift errors. A 
{\it joint} analysis of the two, however, is more robust to these 
uncertainties and leads to a remarkable improvement over the results 
of either probe alone.

Forecasts of cosmological constraints with baryon oscillations and 
weak lensing are provided for four proposed multiband imaging surveys 
in combination with measurements of the cosmic microwave background
from {\it Planck}. In particular, we find that the joint analysis of
galaxy and shear power spectra with the Large Synoptic Survey 
Telescope can tighten the 1$\sigma$ error bounds on the dark energy 
equation of state (at the pivot expansion factor $a_{\rm p} = 0.63$)
%and its rate of change to $0.016$ and $0.16$, respectively. These 
%results are marginalized over 9 other cosmological parameters 
%including the curvature and over 120 parameters that model the 
%linear galaxy clustering bias, photometric redshift bias, and rms 
%photometric redshift error. Simple additive systematic errors are 
%included as nuisance parameters in our forecasts for both optimistic 
%and pessimistic cases. 
and its rate of change, respectively, to $0.016$ and $0.16$ 
(marginalized over 131 other parameters).
With supernovae and cluster counting as well as higher-order 
statistics of the same galaxy and shear data, one can further 
improve the constraints. 

To allow greater flexibility in utilising our forecasts, 
we will make our forecasting tool available publicly.

\noindent{\it Keywords\/}: cosmological constant experiments, 
gravitational lensing, power spectrum, surveys galaxies
\end{abstract}

%\pacs{95.36.+x, 98.62.Py, 98.80.-k}

\section{Introduction}

The apparent accelerated expansion of the universe revealed by
type Ia supernova (SN) distances 
\cite{riess98,perlmutter99a,tonry03,riess04} 
suggests the existence of an unknown component --
dark energy -- of the universe that drives the acceleration with its 
negative equation of state (EOS). Observations of the large-scale 
structure and cosmic microwave background (CMB) also point to
the scenario that the universe is nearly flat with more 
than $70\%$ of its content in dark energy
\cite{dodelson00,percival02,spergel03,tegmark04b,spergel06}. 

Since it is not directly observable, one has to infer dark energy
properties from its impact on the distance--redshift and 
growth--redshift relationship. Aside from SNe, baryon acoustic
oscillations (BAOs) in the galaxy power spectrum and weak 
lensing (WL) can be used to measure the angular diameter distance. 
The BAO technique relies on the 
standard ruler of the sound horizon at the last scattering surface
\cite{peebles70, bond84, holtzman89}, which has been 
accurately measured by the Wilkinson Microwave Anisotropy Probe 
({\it WMAP}) \cite{spergel06} and will be further improved by 
future CMB experiments.
WL measures the distance from its geometric lensing kernel as
well as the shape of shear power spectra, and it is 
sensitive to the growth--redshift relationship \cite{hu99a,hu02b,
huterer02,refregier03,ishak04a,takada04b,song04,knox06b,knox06c}.
Cluster counting is another useful technique that exploit the 
volume--redshift relationship and evolution of cluster number 
density \cite[e.g.][]{haiman01,hu03a}, although it requires a very 
good understanding of the distribution of the mass--observable 
relation including its scatter \cite{lima04,lima05} {\it and} 
the dispersion of the cluster mass function.

The prospects for measuring the dark energy EOS parameters with 
BAO and WL have been studied in great detail 
\cite{eisenstein98,hu99a,cooray01b,hu02b,huterer02,bartelmann01,
blake03,hu03b,linder03,refregier03,seo03,hu04b,ishak04a,song04,
takada04b,angulo05,blake05,glazebrook05,knox06b,seo05,white05,
bernstein06,dolney06,eisenstein06a,eisenstein06b,knox06c,wang06,
zhan06c}. In the absence
of systematic errors and uncertainties of photometric redshift 
(\phz{}) error distribution, a deep and wide multiband WL
survey can achieve statistical precisions of $\sim 0.1$ on 
the dark energy EOS parameters $w_0$ and $w_a$, where the EOS 
is parametrized as $w = w_0 + w_a(1-a)$ and $a$ is the scale factor 
\cite{chevallier01}. Shear 
systematics and uncertainties of \phz{} error distribution can 
severely degrade WL constraints 
\cite{bernstein04,ishak05,ma06,huterer06}. In addition, the lack
of complete understanding of the matter power spectrum on small 
scales due to nonlinearity and baryonic effects 
\cite{white04,zhan04c,hagan05,heitmann05,zhan06a} 
makes it more difficult to fully harness the statistical 
power of the shear power spectrum on those scales \cite{huterer05a}.

Given a deep \phz{} survey, BAO constraints on $w_0$ and $w_a$ are
generally weaker than those of WL, because the rms \phz{} error
$\sigma_z$ filters out radial information on small scales and 
because we do not have an accurate description of the galaxy 
clustering bias \cite{linder03,seo03,glazebrook05,zhan06c,knox06c}. 
When only the auto power spectra are considered, BAO results 
are sensitive to the uncertainties of the \phz{} error 
distribution \cite{dolney06,zhan06c}. However, the 
cross power spectra between redshift bins, as we discuss
below, can self-calibrate the \phz{} bias $\delta z$ and rms 
error $\sigma_z$. As such, tomographic measurements of BAOs
are much less sensitive to the uncertainties of the \phz{} error
distribution.

Although we treat dark energy phenomenologically with a 
loose connection to the cosmological constant or quintessence 
\cite{ratra88,huey99,perlmutter99b,wang00}, one should 
bear in mind that dark energy could also be a manifestation of 
our incomplete knowledge of gravity. For instance,
gravity in an extra dimension can accelerate the cosmic 
expansion \cite{dvali00,deffayet01}. Matching different 
theories to the same 
distance--redshift relationship does not guarantee a match of 
the growth--redshift relationship 
\cite{abazajian03,linder04,simpson05,song05,zhang05,knox06b}. Hence, 
it is important to employ multiple techniques, especially WL, to not 
only strengthen constraints on dark energy properties but also 
potentially distinguish between different theories.

We focus on incorporating the two emerging techniques BAO and WL
with CMB measurements. Both BAOs and cosmic shear have 
been recently detected and used to constrain cosmological parameters 
\cite{bacon00,wittman00,vanwaerbeke01,jarvis03,hoekstra05,
eisenstein05,cole05,huetsi06}. More encouragingly, it has been 
demonstrated that one can achieve an rms \phz{} error 
$\sigma_z \lesssim 0.04(1+z)$ with deep photometries and sufficient
spectroscopic calibrations \cite{ilbert06}.
With a homogeneous population of luminous red galaxies (LRGs),
the rms \phz{} error can be even smaller, and the error distribution
is fitted well by two Gaussians centred at nearly the same true 
redshift \cite{padmanabhan05}. This has led 
to very recent detections of BAO signatures from \phz{} LRG samples 
in the Sloan Digital Sky Survey (SDSS) \cite{padmanabhan06} 
and MegaZ-LRG catalogue \cite{blake06b} at comparable statistical 
significance levels as the detections from spectroscopic samples 
in SDSS \cite{eisenstein05} and the 2 degree 
field galaxy redshift survey \cite{cole05}.

To combine BAO with WL in a single survey, one has to properly 
account for the galaxy--shear correlation 
\cite{schneider98}. We further develop
the work in \cite{hu04b} to include the uncertainties of
the \phz{} error distribution and simple additive systematics. 
\Phz{} errors, even if their 
distribution is known perfectly,
degrade cosmological constraints by suppressing the
signal. For WL, the lensing kernel itself is much broader than the
rms \phz{} error $\sigma_z$, so that it is not overly sensitive to 
\phz{} errors. However, the uncertainties of the \phz{} error 
distribution introduce uncertainties to the shear signal. Given that 
future WL surveys can measure shear power spectra 
to percent level at $\ell \sim 1000$, WL results will be sensitive 
to the uncertainties of the \phz{} error distribution. 
For BAO, the situation is the opposite. Its kernel is defined by the 
\phz{} error distribution (assuming that the widths of the tomographic 
bins match the rms \phz{} errors), so that BAO power spectra and 
their cosmological constraints are more sensitive to \phz{} errors. 
But the distinctive dependency of the tomographic galaxy power 
spectra on \phz{} errors is actually an advantage that can be taken
to constrain the \phz{} error distribution. Consequently, BAO
results are less prone to the uncertainties in the \phz{} error
distribution. 

\Phz{} errors can have long tails and sometimes can be
catastrophically wrong due to the confusion of spectral features 
(especially the Balmer break for galaxies at $z\lesssim 1$ and
the Lyman break for galaxies at $z \gtrsim 2$) or the lack of 
features in the wavebands \cite[for a review, see][]{yee98}.
Such catastrophic errors can be mitigated by incorporating 
galaxy apparent magnitude and size priors or by supplementing 
near-infrared data \cite{connolly97,benitez00,fontana00,babbedge04}.
The outliers have been reportedly limited to less than $10\%$ or 
even completely eliminated in some cases 
\cite{benitez00,babbedge04,mobasher04,ilbert06}.

We idealise the \phz{} errors to be Gaussian, so that they can be 
parametrized with the rms $\sigma_z$ and bias $\delta z$. The 
Gaussian simplification allows for a pedagogical understanding of 
the effect of the uncertainties in the \phz{} error distribution. 
The BAO constraints on the \phz{} error distribution will be 
weakened when more parameters are used to 
model the distribution \cite{schneider06}. However, since 
WL cannot self-calibrate the \phz{} parameters, even very loose 
error bounds on these parameters from BAO will keep WL constraints
on dark energy from being degraded without bound. 
Spectroscopic calibrations are crucial for both properly 
modelling the \phz{} errors and constraining their distribution, but,
it will be very challenging to fairly sample galaxies for 
spectroscopy over the entire area and 
depth of future imaging surveys. Therefore, alternative means of 
constraining the \phz{} error distribution, such as \phz{} galaxy 
power spectra \cite{schneider06} and cross-correlations 
between spectroscopic and photometric galaxy samples 
\cite{newman06}, will be valuable to these surveys.

\Phz{} surveys can be used in many ways to probe the universe, and a
single technique can utilise a variety of statistics. We specifically
investigate the complementarity between tomographic galaxy and 
shear power spectra. Higher-order statistics can be very useful 
\cite{takada04b,dolney06,huterer06}, but to properly 
combine them with other statistics, one needs to carefully account 
for the correlations between different statistics. 

The rest of this paper is organised as follows. We define the 
tomographic galaxy and shear power spectra as well as the CMB power 
spectra in section \ref{sec:obs}, and discuss the Fisher information 
matrix for error estimation in section \ref{sec:est}. Our forecasts
on cosmological constraints are given in section \ref{sec:resl} for 
four proposed multiband imaging surveys (as far as BAO and WL are
concerned): 
Dark Energy Survey\footnote{See http://www.darkenergysurvey.org/.} 
(DES), Large Synoptic Survey Telescope\footnote{See 
http://www.lsst.org/.} (LSST), Panoramic Survey Telescope and
Rapid Response System\footnote{See 
http://pan-starrs.ifa.hawaii.edu/.} (PS4), and 
Supernova/Acceleration Probe\footnote{See http://www.snap.lbl.gov/.
As its name suggests, SNAP is also designed to measure SN 
redshifts with spectroscopy.}
(SNAP). We discuss the results and conclude in section \ref{sec:dis}.

\section{Observables} \label{sec:obs}

Our observables consist of angular maps of galaxy number density
$n(\btheta)$, shear $\gamma(\btheta)$, and CMB temperature 
and polarisation fluctuations. The tomographic galaxy and shear power 
spectra are the covariance of galaxy number density and shear in 
Fourier space, as are the CMB temperature and polarisation power 
spectra to temperature and polarisation. In what follows, we loosely 
refer to these power spectra as observables as well, despite that 
using power spectra as true observables in the forecast 
requires a slightly different formalism.

\subsection{Galaxy and shear power spectra}
With the Limber approximation \cite{limber54,kaiser92}, the angular
power spectrum of $n(\btheta)$ and $\gamma(\btheta)$ can be written 
as \cite{hu04b}
\be \label{eq:aps}
P_{ij}^{XY}(\ell) = \frac{2\pi^2}{c\ell^3} \int \rmd z\, H(z) 
D_{\rm A}(z) W_i^X(z) W_j^Y(z) \Delta^2_\delta(k; z),
\ee
where lower case subscripts correspond to the tomographic bins, upper 
case superscripts label the observables, i.e., $X=$ g for galaxies 
or $\gamma$ for shear, $H(z)$
is the Hubble parameter, $D_{\rm A}(z)$ is the comoving angular diameter 
distance, $\Delta^2_\delta(k;z)$ is the dimensionless power spectrum
of the density field, and $k = \ell/D_{\rm A}(z)$. BAO and WL do not 
necessarily use the same binning. In other words, the bin number is  
defined for each technique separately. The window function is
\be \label{eq:Wi}
W_i^X(z) = \cases{b(z)\frac{n_i(z)}{\bar{n}_i} & $X$ = g \\
\frac{3}{2}\frac{\Omega_{\rm m}H_0^2}{H(z)}\frac{D_{\rm A}(z)}{a\,c} 
\int_z^\infty \!\rmd z'\, \frac{n_i(z')}{\bar{n}_i}
\frac{D_{\rm A}(z,z')}{D_{\rm A}(z')} & $X$ = $\gamma$,}
\ee
where $b(z)$ is the linear galaxy clustering bias, and 
$\Omega_{\rm m}$ and $H_0$ are, respectively, the matter 
fraction at $z = 0$ and Hubble constant. 
The galaxy redshift distribution $n_i(z)$ in the
$i$th tomographic bin is an average of the underlying 
three-dimensional galaxy distribution over angles, and the mean
surface density $\bar{n}_i$ is the total number of galaxies per
steradian in bin $i$. The distribution $n_i(z)$ is broader 
than the nominal width of the tomographic bin (defined in \phz{} 
space) because of \phz{} errors.

We only include galaxy power spectra on largely linear scales, 
e.g., the scales of BAOs, in our analysis, 
so that we can map the matter power spectrum to galaxy power
spectrum with a scale-independent but time-evolving linear 
galaxy bias \cite{verde02,tegmark04a}. 
One may extend the analysis to smaller scales with a halo model 
to describe the scale dependency of the galaxy bias and, in fact, 
can still constrain the scale-dependent galaxy bias to $1\%$ 
level \cite{hu04b}. 

Cast in the form of \eref{eq:aps} and \eref{eq:Wi}, it seems as 
though the WL kernel $W_i^\gamma(z)$ is sensitive to the matter 
fraction $\Omega_{\rm m}$ (or the matter density 
$\omega_{\rm m} = \Omega_{\rm m} h^2$, where $h$ is the
Hubble constant in units of 100 \mbox{km\,s$^{-1}$\,Mpc$^{-1}$}). 
Potential (or spatial curvature) 
perturbations, for which inflation often predicts a nearly 
scale-invariant behaviour, are more fundamental than density 
perturbations. WL shear and CMB temperature fluctuations are also 
driven directly by the fluctuations in the gravitational 
potential rather than the density. Therefore, we recast 
the angular power spectra in terms of the potential power spectrum:
\be \label{eq:apphi}
P_{ij}^{XY}(\ell) = \frac{2\pi^2\ell}{c} \int \rmd z\,  
H D_{\rm A} W_i^X W_j^Y \Delta^2_\phi(k),
\ee
with
\be \label{eq:Wiphi}
W_i^X = \cases{\frac{n_i}{\bar{n}_i} 
\frac{2 a\,b}{3 \Omega_{\rm m} H_0^2 D_{\rm A}^2}
& $X$ = g \\ \frac{1}{c\,H D_{\rm A}}
\int_z^\infty \!\rmd z'\, \frac{n_i(z')}{\bar{n}_i}
\frac{D_{\rm A}(z,z')}{D_{\rm A}(z')} & $X$ = $\gamma$.}
\ee
We have dropped the argument $z$ in \eref{eq:apphi} and 
\eref{eq:Wiphi} where there is no confusion. One may choose to 
normalise to either the density, e.g., the rms density fluctuation
$\sigma_8$ within a radius of $8\,\mpc$  at $z=0$, or the 
potential, e.g., the CMB temperature power spectra, but one should 
propagate the factors of $\omega_{\rm m}$ through to ensure that 
gravitational lensing responds to potential directly.
Moreover, precise measurements of CMB power spectra have made 
it more desirable as well as convenient to normalise to the 
potential fluctuations.

Observationally, the power spectra will have contributions from
galaxy shot (shape) noise $\bar{n}_i^{-1}$ 
($\gamma_{\rm rms}^2 \bar{n}_i^{-1}$) and systematic noise 
$N_{\rm sys}^X$:
\be \label{eq:totps}
\tilde{P}_{ij}^{XY}(\ell) = P_{ij}^{XY}(\ell) +
\delta_{XY}^{\rm K}\left(N_{\rm sys}^X + 
\delta_{ij}^{\rm K}\frac{X_{\rm rms}^2}{\bar{n}_i}\right),
\ee
where $\delta_{XY}^{\rm K}$ and $\delta_{ij}^{\rm K}$ are 
Kronecker delta functions. For galaxies, ${\rm g_{rms}} \equiv 1$, 
and, for the shear, $\gamma_{\rm rms}\sim 0.2$ is due to the 
intrinsic shape of galaxies and measurement errors.
%rev1
The systematic noise $N_{\rm sys}^X$ may arise from (the residuals 
after correcting for) the spatially varying dust extinction, 
photometry offsets, point spread function (PSF), instrumentation 
effects, and so on \cite{tegmark02,guzik05,jain06}.
%end
We have assumed in \eref{eq:totps} that $N_{\rm sys}^X$ is 
additive, uncorrelated between BAO and WL, and independent of 
tomographic bins or multipoles. This is a great simplification over
the additive errors considered in \cite{huterer06}, which shows that
the degradation to WL results by the systematics is generally 
not bound. The systematic noise we include is among a sub-set of 
(unrealistic) additive errors whose degradation effect can be 
shown mathematically to be less than $\sim 100\%$ \cite{huterer06}. 
As such, we have effectively made an aggressive assumption that all 
other types of systematics are well under control.

\begin{figure}
\centering
\includegraphics[width=5in]{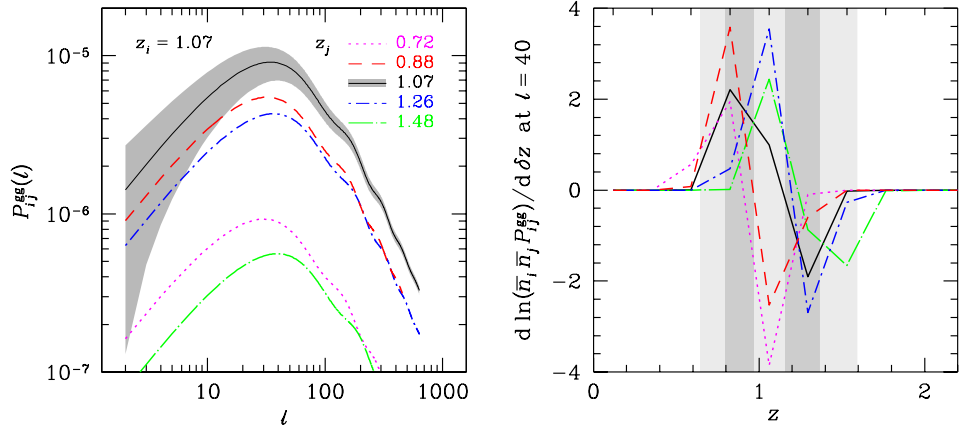}
\caption{{\it Left panel}: Angular galaxy power spectra 
$P_{ij}^{\rm gg}(\ell)$ with bin $i$ centred at $z_{\rm p} = 1.07$ 
and bin $j$ centred at $z_{\rm p} = 0.72$ (dotted line), $0.88$ 
(dashed line), $1.07$ 
(solid line), $1.26$ (dash-dotted line), and $1.48$ 
(long-dash-dotted line). The shaded area indicates 1$\sigma$ 
statistical errors of the auto power spectrum 
$\Delta P_{ii}^{\rm gg}(\ell) = [(\ell+0.5)f_{\rm sky}]^{-1/2} 
\tilde{P}_{ii}^{\rm gg}(\ell)$, where $f_{\rm sky}$ is the sky 
coverage, for LSST. The right end of each power spectrum curve is set 
by an $\ell_{\rm max}$ for each tomographic bin to reduce 
contaminations from nonlinearity. 
{\it Right panel}: Sensitivity of the observables 
$\bar{n}_i \bar{n}_j P_{ij}^{\rm gg}(\ell)$ to the 
\phz{} bias $\delta z$. The derivatives are evaluated at 
$\ell = 40$ to avoid contributions from baryon oscillations that 
are prominent at $\ell \gtrsim 100$. Though not shown, 
$\bar{n}_i \bar{n}_j P_{ij}^{\rm gg}(\ell)$ is also sensitive 
to the \phz{} rms error. The grey 
vertical bands correspond to the relevant tomographic bins in \phz{} 
space, while the upper ticks mark the positions of \phz{} 
parameters in true redshift space. For demonstration purpose, we
use 15 tomographic bins from $z_{\rm p} = 0.15$ to $3.5$ in \phz{} 
space and 17 \phz{} parameters uniformly distributed from $z = 0$ to 
$4$ in true redshift space. The widths of the bins are 
proportional to $(1+z_{\rm p})$.
\label{fig:cldcl}}
\end{figure}

Previous forecasts for BAO use only the three or two-dimensional 
galaxy auto power spectrum within each redshift bin \cite{blake03,
linder03,seo03,blake05,glazebrook05,dolney06,wang06,zhan06c}. This 
is a good approximation for spectroscopic surveys, because the 
intrinsic correlation (due to the large-scale structure) between two 
redshift bins is small if the bins are wide enough. 
For \phz{} surveys, the cross-bin power 
spectra $P_{ij}^{\rm gg}(\ell)$ are no longer negligible due to 
\phz{} errors. The left panel of \fref{fig:cldcl} shows that 
$P_{ij}^{\rm gg}(\ell)$ (broken lines), depending on the binning 
and \phz{} errors, can have comparable amplitudes to that of the 
auto power spectrum $P_{ii}^{\rm gg}(\ell)$ (solid line).
Not including the covariance between observables in different
bins, i.e., the cross power spectra, generally leads to more 
optimistic error estimates. 

Aside from statistical necessity, tomographic galaxy power spectra 
will also be very useful for parameter estimation with \phz{} BAO 
experiments, because they provide valuable information about \phz{} 
errors. The 
reason is that, with Limber approximation, the cross power spectra 
$P_{ij}^{\rm gg}(\ell)$ are determined by the overlap of galaxy 
number density distributions $n_i$ and $n_j$ in true redshift 
space and thus sensitive to \phz{} bias $\delta z$ and rms error 
$\sigma_z$. This is demonstrated in the right panel of 
\fref{fig:cldcl}, where derivatives of the new 
observables\footnote{It is argued in \cite{ma06} that 
$\bar{n}_i \bar{n}_j P_{ij}^{\gamma\gamma}(\ell)$ is 
the fundamental observable for WL. Our forecasts for cosmological 
parameters are not affected by this change of observables, but the 
constraints on \phz{} parameters are slightly improved.} 
$\bar{n}_i \bar{n}_j P_{ij}^{\rm gg}(\ell)$ with respective to 
\phz{} bias parameters (see section \ref{sec:est} for details) are 
given at $\ell = 40$. The extents of the relevant tomographic bins 
are indicated by grey vertical bands, and the locations of the 
\phz{} parameters are 
marked by upper ticks. The \phz{} bias at an arbitrary redshift is 
linearly interpolated between two \phz{} bias parameters.

We refer to the galaxy bin centred at $z_{\rm p} = 1.07$ (the 
subscript p following $z$ signifies \phz{}s) as bin $i$ in the
following discussion. An increase of the \phz{} bias $\delta z$ 
at $z = 0.82$ means
that the redshifts of galaxies around $z = 0.82$ are more likely to 
be overestimated. As a result, more of these galaxies will be 
assigned to bin $i$. With no change to the contributions of galaxies 
from other redshifts, the auto power spectrum 
$\bar{n}_i^2 P_{ii}^{\rm gg}$ (solid line) increases correspondingly.
The product of the fiducial galaxy clustering bias and linear
growth function evolves less than $1\%$ in this redshift range, so 
that the change in the mixing of galaxy number fluctuations at 
different redshift is sub-dominant to the effect of increased 
$\bar{n}_i$. Similarly, an increase of $\delta z$ at $z = 1.29$ 
leads to fewer $z \sim 1.29$ galaxies assigned to bin $i$ and 
thus lowers $\bar{n}_i^2 P_{ii}^{\rm gg}$. 

The behaviour of the auto power spectrum $\bar{n}_i^2 P_{ii}^{\rm gg}$
as a function of the \phz{} bias $\delta z$ near the nominal centre of 
bin $i$ depends on the overall galaxy distribution $n(z)$. 
An increase of $\delta z$ at $z = 1.06$ 
(by linear interpolation, it increases the 
\phz{} bias in the range $0.82 < z < 1.29$) 
will shift some galaxies from bin $i-1$ to bin $i$ and some from
bin $i$ to bin $i+1$. Since the fiducial $n(z)$ peaks at $z = 1$, 
the net result is more galaxies in bin $i$ and, hence, a positive 
response of $\bar{n}_i^2 P_{ii}^{\rm gg}$ to $\delta z$ at $z = 1.06$. 

The cross power spectra are more sensitive to the tail 
of the \phz{} error distribution, because they are determined by 
the overlap between two tomographic bins in true redshift space. 
Although the increase of $\delta z$ at $z = 0.82$ leads to fewer 
galaxies in lower-redshift bins (dotted line and dashed line), 
the increased overlap between these bins and bin $i$, i.e., 
bin $i$ getting more $z \sim 0.82$ galaxies, results in a 
boost to $\bar{n}_i \bar{n}_j P_{ij}^{\rm gg}$ for $j < i$. 
The increase of $\delta z$ at $z = 1.06$ reduces the number of
$z \sim 1.06$ galaxies assigned to the lower-redshift bins, so that 
$\bar{n}_i \bar{n}_j P_{ij}^{\rm gg}$ decreases.
The behaviour of the cross power spectra between higher redshift 
bins and bin $i$ (dash-dotted line and long-dash-dotted line) can 
be explained similarly.

For our forecasts, we use {\sc cmbfast} version 4.5.1 
\cite{zaldarriaga00} 
to calculate the matter transfer function at $z = 0$, and then apply 
the linear growth function and Peacock \& Dodds fitting formula 
(PD96) \cite{peacock96} to obtain the nonlinear matter power 
spectrum at any redshift. Details are given in \ref{app:ling} and
\ref{app:nlps}. To reduce contaminations of the small-scale 
nonlinearity and baryonic effects 
\cite{seo03,white04,zhan04c,hagan05,heitmann05,seo05,white05,
zhan06a}, we set a smallest angular scale in our analysis. 
Separately, to avoid possible effects of dark energy clustering
that have not been considered in our analysis, 
we also discard galaxy and 
shear statistics on very large angular scales \cite{song04}. This
limits the multipoles $40 \le \ell \le 2000$ for WL. The BAO kernel
is rather narrow, so it is more appropriate to set the criterion in
$k$ space. We require that the dimensionless power spectrum 
$\Delta_\delta^2(k;z) < 0.4$ in each tomographic bin \emph{and} 
that $40 \le \ell \le 3000$ for BAO. 
%rev1
The maximum multipole for a galaxy bin centred at $z_{\rm p}$ 
follows approximately 
$\ell_{\rm max} =340z_{\rm p} + 346z_{\rm p}^2$ at
$0.15 \le z_{\rm p} \le 0.9$, 
$446 - 658z_{\rm p} + 908z_{\rm p}^2$ at 
$0.9 < z_{\rm p} \le 2.1$, and 3000 at $z_{\rm p} > 2.1$.
%\be
%\ell_{\rm max} =\cases{
%340z_{\rm p} + 346z_{\rm p}^2 & $0.15 \le z_{\rm p} \le 0.9$ \\
%446 - 658z_{\rm p} + 908z_{\rm p}^2 & $0.9 < z_{\rm p} \le 2.1$ \\
%3000 & $z_{\rm p} > 2.1$. }
%\ee
%end
Setting the minimum multipole to 100
or reducing the maximum multipole by 1000 for both BAO and WL
does not severely impact our forecasts (see section \ref{sec:dis}).

\subsection{Cosmic microwave background}

CMB temperature and polarisation power spectra are calculated with
{\sc cmbfast} for scalar modes:
\be
\mathcal{C}_{\ell}^{XY} = 2(2\pi)^4\int \rmd \ln k\,
\Delta_\phi^2(k) \Delta_{\ell}^X(k) \Delta_{\ell}^Y(k),
\ee
where $X=$ T for temperature or E for E-polarisation, and
$\Delta_{\ell}^X(k)$ is the radiation transfer function 
\cite{zaldarriaga97}. With noise, the power spectra become 
\be
\tilde{\mathcal{C}}_{\ell}^{XY} = \mathcal{C}_{\ell}^{XY} + 
\delta_{XY}^{\rm K}
\left(\sum_i \theta_{{\rm b},i}^{-2} X_{{\rm rms},i}^{-2} 
B_{{\ell},i}^2\right)^{-1},
\ee
where $\theta_{{\rm b},i}$ is the full width half-maximum of the 
beam in channel $i$, $X_{{\rm rms},i}$ is the noise per pixel in 
temperature or polarisation measurements,
and $B_{{\ell},i}$ is the beam window function
\cite{zaldarriaga97,eisenstein99b}. We model the CMB
experiment after {\it Planck} with instrumentational 
characteristics tabulated in \cite{kaplinghat03}. We assume a 
sky fraction of $f_{\rm sky} = 0.8$ after a foreground cut, and limit 
the multipoles $2 \le \ell \le 2000$ for temperature
and $10 \le \ell \le 2500$ for polarisation.

CMB temperature and polarisation are nearly uncorrelated with 
the galaxy distribution except for the late-time integrated 
Sachs-Wolfe (ISW) effect on large angular scales \cite{sacks67}. 
Since we do not use BAO or WL at $\ell < 40$, it is 
straightforward to combine CMB with BAO and WL. 

\section{Error estimation} \label{sec:est}

Assuming that the likelihood function of the true observables, e.g., 
the map data, is approximately a multivariate Gaussian around the 
maximum, one can use the Fisher information matrix to estimate the 
lower bound of errors of the parameters that could be inferred 
\cite{jungman96,vogeley96,tegmark97a,tegmark97b}.

\subsection{Fisher matrix}

The Fisher matrix of parameters $\{p_\alpha\}$ is given by 
\cite{tegmark97b}
\be \label{eq:fullfish}
F_{\alpha\beta} = \frac{1}{2} \Tr 
\bi{C}^{-1} \frac{\partial \bi{C}}{\partial p_\alpha} 
\bi{C}^{-1} \frac{\partial \bi{C}}{\partial p_\beta} +
\frac{\partial \bmu^{\rm T}}{\partial p_\alpha} \bi{C}^{-1} 
\frac{\partial \bmu}{\partial p_\beta},
\ee
where $\bi{C}$ and $\bmu$ are, respectively, the covariance and mean
of the true observables. For our purpose, the observables have 
zero mean, and the contributions to the Fisher matrix from different
spherical harmonic modes are separable, so that
\be \label{eq:trfish}
F_{\alpha\beta} = f_{\rm sky} \sum_\ell \frac{2\ell + 1}{2} \Tr 
\bi{C}_\ell^{-1} \frac{\partial \bi{C}_\ell}{\partial p_\alpha} 
\bi{C}_\ell^{-1} \frac{\partial \bi{C}_\ell}{\partial p_\beta},
\ee
with $(\bi{C}_\ell)_{ij}^{XY} = \tilde{P}_{ij}^{XY}(\ell)$
for galaxies and shear (BAO+WL), and 
$(\bi{C}_\ell)^{UV} = \tilde{\mathcal{C}}_\ell^{UV}$ for CMB. The 
factor $f_{\rm sky}$ approximates the effect of incomplete sky 
coverage with a reduction of modes. This is valid for a contiguous 
survey with roughly the same angular size in all directions. 
The total Fisher matrix is a direct sum of the BAO+WL and CMB Fisher 
matrices.

\Eref{eq:trfish} can be reduced to
\be \label{eq:fish}
F_{\alpha\beta} = f_{\rm sky} \sum_\ell (2\ell + 1)
\frac{\partial \bi{Q}_\ell^{\rm T}}{\partial p_\alpha} 
\bi{M}_\ell^{-1} 
\frac{\partial \bi{Q}_\ell}{\partial p_\beta},
\ee
where $\bi{Q}_\ell$ is a column vector containing unique
elements of $\bi{C}_\ell$, and $\bi{M}_\ell$ equals the 
covariance of $\bi{Q}_\ell$ in the Gaussian case \cite{hu04b}.
From \eref{eq:fish}, it may seem possible to discard the cross 
power spectra in error forecasts. However, the Fisher matrix 
\eref{eq:fullfish} and \eref{eq:fish} are based on the multivariate 
Gaussian likelihood of the true observables, e.g.,
\be
-2\ln \mathcal{L}(\bi{n}_\ell) = \ln |\bi{C}_\ell^{\rm gg}| + 
\bi{n}_\ell^{\rm T} (\bi{C}_\ell^{\rm gg})^{-1} 
\bi{n}_\ell + \mbox{const},
\ee
where $\bi{n}_\ell$ is a column vector of the galaxy angular 
distribution in multipole space, whereas the likelihood becomes
\be
-2\ln \mathcal{L}'(\bi{n}_\ell) = 
\sum_i \left[\ln C_{ii, \ell}^{\rm gg} + 
n_{i, \ell}^2 (C_{ii, \ell}^{\rm gg})^{-1}\right] + {\rm const},
\ee
if the cross power spectra are discarded. The two likelihood 
functions $\mathcal{L}$ and $\mathcal{L}'$ can differ considerably, 
%rev1
when, for instance, \phz{} errors induce strong correlations 
of the galaxy number density fluctuations between the bins.
Discarding the correlations between the observables may
lead to underestimation of the errors on cosmological parameters
\cite[e.g.,][]{cooray01a}.
However, in the case of BAO, the galaxy cross power spectra are
crucial for calibrating the \phz{} error distribution, which is, 
in turn, critical to measuring the dark energy EOS parameters
accurately with the WL technique. Consequently, the joint BAO and 
WL constraints on dark energy can be weakened if one does not 
utilise the galaxy cross power spectra (see the footnote of 
\tref{tab:opt}).
%end

To use the power spectra as true observables, one can replace
$\bmu$ and $\bi{C}$ in \eref{eq:fullfish} with the power spectra and
their covariance, respectively. In doing so, one also assumes that
the likelihood function of the power spectra is a multivariate 
Gaussian, which is generally not the same as assuming a Gaussian 
likelihood function for the map data.

\subsection{Parameters}

Our parameter set includes 11 cosmological parameters, 40 galaxy
bias parameters, 40 \phz{} bias parameters, 40 \phz{} rms 
parameters, one systematic noise parameter for BAO, and one
systematic noise parameter for WL.
%rev1
There are 133 parameters in total for the joint BAO and WL analysis. 
When the BAO and WL techniques are considered separately, their 
respective irrelevant parameters are held fixed.
%end

The cosmological parameters are the dark energy EOS parameters
$w_0$ and $w_a$, the matter 
density $\omega_{\rm m}$, the baryon density $\omega_{\rm b}$, the 
angular size of the sound horizon at the last scattering surface 
$\theta_{\rm s}$, the equivalent matter fraction of curvature 
$\Omega_{\rm K}$, the optical depth $\tau$ to scattering by electrons 
in the reionized intergalactic medium, the primordial helium mass
fraction $Y_{\rm p}$, the spectral index $n_{\rm s}$ of the primordial 
scalar perturbation power spectrum, the running of the spectral index 
$\alpha$, and the normalisation of the primordial curvature power 
spectrum $\Delta_R^2$ (or $\case{25}{9}\Delta_\phi^2$ in matter era) 
at $k = 0.05\,\mbox{Mpc}^{-1}$. We adopt the three-year {\it WMAP} 
results \cite{spergel06} for fiducial values of the parameters: 
($w_0$, $w_a$, $\omega_{\rm m}$, $\omega_{\rm b}$, $\theta_{\rm s}$, 
$\Omega_{\rm K}$, $\tau$, $Y_{\rm p}$, $n_{\rm s}$, $\alpha$, 
$\Delta_R^2$) =
($-1$, 0, 0.127, 0.0223, 0.596$^\circ$, 0, 0.09, 0.24, 0.951, 0,  
$2.0\times 10^{-9}$). The reduced Hubble constant $h = 0.73$ and the
present equivalent matter fraction of dark energy 
$\Omega_{\rm X} = 0.76$ are implicit in our parametrization. 
 
The \phz{} bias and rms error parameters are implemented 
according to \cite{ma06}; we evenly space the parameters from $z = 0$
to $4$ and linearly interpolate the values at arbitrary redshifts.
%rev1
Note that the \phz{} parameters are assigned in true-redshift space
and that they are not attached to any galaxy or shear bins.
%end
The galaxy clustering bias is parametrized in the same way. 
This scheme is based on the expectation that \phz{}s and the galaxy 
bias behave more or less smoothly from one redshift to 
another. 
%rev1
It is found that the WL forecasts for a DES-like survey become 
roughly convergent when the redshift interval of the \phz{} 
parameters is less than 0.15 \cite{ma06}. 
The convergent redshift interval may be different for other 
surveys, but it will be very expensive to achieve the same 
precision of the \phz{} calibration via spectroscopy at a much 
finer interval.
%end
We expect that ambitious \phz{} calibrations will be 
carried out at $\sim 0.1$ redshift intervals 
(LSST \phz{} calibration white 
paper\footnote{See http://www.lsst.org/Science/Phot-z-plan.pdf.}) 
and that similar measurements of 
the galaxy clustering bias will be available. 
%rev1
This amounts to 40 parameters each for the galaxy bias, \phz{}
bias, and \phz{} rms over $0 \le z \le 4$.
%end
The actual calibration
requirement may be less stringent if one combines BAO with WL
(see section \ref{sec:resl}).
The \phz{} and galaxy clustering bias parameters extend beyond 
the nominal cut-off (photometric) redshift of the surveys to 
accommodate the galaxy distribution of the last tomographic bin in 
true redshift space. Our forecasts are not affected by this 
extension, because the low galaxy number density and poor 
measurements reduce the statistical significance of the data around
the cut-off redshift.

We assume $b = 1 + 0.84\,z$ \cite{weinberg04}, $\delta z = 0$, and 
$\sigma_z \propto (1+z)$ for the fiducial model. The galaxy 
distribution $n_i$ in each tomographic bin is determined from 
the overall galaxy redshift distribution $n(z)$ and \phz{} 
parameters
\be
n_i = n(z) P(z_{{\rm p},i}^{\rm B}, z_{{\rm p},i}^{\rm E}; z)
= n(z) I(z_{{\rm p},i}^{\rm B}, z_{{\rm p},i}^{\rm E}; z) /
I(0,\infty; z),
\ee
where $z_{{\rm p},i}^{\rm B}$ and $z_{{\rm p},i}^{\rm E}$ define the 
extent of bin $i$, and 
$P(z_{{\rm p},i}^{\rm B}, z_{{\rm p},i}^{\rm E}; z)$ is the 
probability of assigning a galaxy that is at true redshift $z$ to 
bin $i$. The unnormalised probability $I(a,b;z)$ is given by
\cite{ma06}
\be
I(a, b; z) = \frac{1}{\sqrt{2\pi}\,\sigma_z} \int_a^b \rmd z_{\rm p} 
\,\exp\left[-\frac{(z_{\rm p} - z - \delta z)^2}{2\sigma_z^2}\right],
\ee
and the normalisation $I(0,\infty,z)$ is enforced because \phz{}s are 
non-negative. Note that even though the probability distribution of
\phz{}s at a given true redshift is assumed Gaussian, the reverse is
not true. In other words, the Gaussian assumption is flexible enough
to allow for modelling of more complex galaxy distributions in
tomographic bins \cite{ma06}.

\begin{table}
\caption{Priors and additive systematics$^{\,\rm a}$.
\label{tab:case}}
\lineup
\begin{indented}
\item[]\begin{tabular}{@{}lcccc}
\br
Case & $\sigma_{\rm P}(\delta z)^{\,\rm b}$  & 
$\sigma_{\rm P}(\ln b)$ & 
$N_{\rm sys}^{\rm g}$ & $N_{\rm sys}^{\gamma}$ \\
\mr
Opt. & 0.05$\,\sigma_z$ & 0.15 & $10^{-8}$ & $10^{-8}$ \\
Pes. & 0.2\0$\,\sigma_z$ & 0.3\0 & $2\times 10^{-7}$ & 
$2\times 10^{-7}$ \\
\br
\end{tabular}
\item[] $^{\rm a}$ The additive systematic errors are defined 
in \eref{eq:totps}.
\item[] $^{\rm b}$ $\sigma_{\rm P}(\sigma_z) = 
\sqrt{2} \sigma_{\rm P}(\delta z)$.
\end{indented}
\end{table}

The derivatives in \eref{eq:fish} are taken two-sided numerically 
with steps in parameters
$\Delta  (w_0, w_a, \omega_{\rm m}, \omega_{\rm b}, \theta_{\rm s}, 
\Omega_{\rm K}, \tau, Y_{\rm p}, n_{\rm s}, \alpha) = \pm 10^{-3}
\times (20, 40, 6, 1, 3, 10, 10, 20, 5, 5)$. The derivative
with respect to $\Delta_R^2$ is analytic. The steps of the galaxy 
clustering bias parameters are $\pm 0.005 b$, and those of the 
\phz{} parameters are $\pm 0.005(1+z)$. When differentiating with
respect to the \phz{} parameters, we keep the 
overall underlying distribution of galaxies, i.e., $n(z)$, 
invariant, because the true redshift of a galaxy cannot be 
affected by \phz{} algorithms. This approach is complementary to 
that of \cite{dolney06,huterer06,schneider06,zhan06c}, in which
the number of galaxies in each tomographic bin is fixed or tightly 
constrained while $n(z)$ is allowed to vary. 
The two approaches are found to be 
consistent for WL \cite{huterer06}. For BAO, $n(z)$ will 
be largely degenerate with the galaxy clustering bias.
%[see \eref{eq:Wi}]. 
Given that our BAO results on the galaxy bias are roughly
consistent with those in \cite{dolney06}, which achieve 
$\sim 2\%$ constraints for a $z < 1.3$ and $f_{\rm sky}=0.1$ 
survey with $\sigma_{\rm P}(\delta z) = 0.01(1+z)$, 
the two approaches can be consistent for BAO as well. 

We treat the additive systematics of BAO and WL as nuisance 
parameters. Since they can be determined very well from the data, 
the additive systematics affect the constraints mostly by boosting 
the covariance \cite{huterer06}. Our forecasts will not improve
even if $N_{\rm sys}^{\rm g}$ and $N_{\rm sys}^\gamma$ are fixed.
This is somewhat misleading, 
because even a very slight modification to the form of the 
systematics can lead to unrestrained degradation to cosmological
constraints for WL \cite{huterer06}. Therefore, one must 
adequately calibrate and control all sorts of systematics in 
order to achieve the constraints we forecast.

We incorporate independent measurements of galaxy bias and \phz{} 
parameters as Gaussian priors in the Fisher matrix. Since the 
results depend on these priors and the systematics to some degree, 
we list our assumptions in \tref{tab:case} for both pessimistic and 
optimistic cases. Note that we use a subscript P for priors, i.e., 
$\sigma_{\rm P}(p_\alpha)$, to distinguish them from constraints
$\sigma(p_\alpha)$. For Gaussian \phz{} errors, a prior of 
$\sigma_{\rm P}(\delta z) = 0.05 \sigma_z$ translates to 400 fairly
sampled spectra for calibration around the redshift of the \phz{} 
bias parameter. Current determination of the galaxy clustering bias
is at $10\%$ level for low redshift galaxies
\cite{hoekstra02,verde02,seljak05}. 
This may or may not be achievable at higher redshift, but it 
is not critical as BAO and WL, along with CMB, can determine 
the galaxy clustering bias quite well. 

%rev1
We infer from the systematic errors (including
extinction, photometry calibration, and seeing) of the SDSS 
angular galaxy power spectrum \cite{tegmark02} that 
$10^{-8} \lesssim N_{\rm sys}^{\rm g} \lesssim$ a few $10^{-7}$. 
For WL, the specified systematic shear noise is somewhat arbitrary 
and perhaps conservative, because its amplitude is relatively high 
compared to that of the shear power spectra. This partially
compensates the fact that we restrict $N_{\rm sys}^\gamma$ 
to a very simple additive form. We also note that the effect of 
the simple noise $N_{\rm sys}^\gamma$ saturates quickly at 
$N_{\rm sys}^{\gamma} \gtrsim$ a few $10^{-8}$ \cite{huterer06}.
In what follows, we assume the optimistic case by default, and the
pessimistic results are presented in \tref{tab:pes}.
%end

\section{Forecasts} \label{sec:resl}

The puzzle of dark energy has inspired a number of ambitious 
photometric and spectroscopic surveys that are proposed to 
unravel the mystery via multiple techniques. 
Among them, we choose to forecast cosmological constraints of BAO 
and WL for four multiband imaging surveys: DES, LSST, PS4, and SNAP
in conjunction with CMB measurements by {\it Planck}, which is 
always included. 
It should be noted that all the four surveys 
propose to utilise one or two other techniques, 
SN and cluster counting, to constrain cosmology, and that BAO 
and/or WL may not even be their primary dark energy probe. 

We use the error product (EP) 
\be \label{eq:ep}
\sigma(w_{\rm p})\times\sigma(w_a) =
[\sigma^2(w_0) \sigma^2(w_a) - {\rm Cov}^2(w_0, w_a)]^{1/2},
\ee 
where ${\rm Cov}(w_0, w_a)$ is the covariance between $w_0$ and
$w_a$, to assess dark energy constraints throughout this section.
The minimum error of the dark energy EOS $\sigma(w_{\rm p})$ is 
achieved at the pivot expansion factor
\be \label{eq:pivot}
a_{\rm p} = 1 + {\rm Cov}(w_0, w_a) / \sigma^2(w_a) 
\ee
and is equal to the error on $w_0$ with $w_a$ held fixed 
\cite{eisenstein99b,hu04b,martin06}.
This product is proportional to the 
area of the error ellipse in the $w_0$--$w_a$ plane and inversely
proportional to the figure of merit used by the Dark Energy 
Task Force.

\begin{table}
\caption{Survey specifications.
\label{tab:specs}}
\lineup
\begin{indented}
\item[]\begin{tabular}{@{}lcccccc}
\br
& Area & & Depth & $\bar{n}$ & & $\sigma_z$ \\
Survey & deg$^2$ & Filters & $r$ mag $(10\,\sigma)$ & 
arcmin$^{-2}$ & z$^*$ & $(1+z)$\\
\mr
DES  & \05000 & {\it griz} & 24.1 & \0\08 & 0.3 & 0.07 \\
LSST &  20000 & {\it ugrizy} & 26.5 & \050 & 0.5 & 0.05 \\
PS4  &  20000 & {\it grizy} & 24.1 & \0\08 & 0.3 & 0.07 \\
SNAP & \01000 & 9 bands$^{\rm a}$ & 26.6 & 100 & 0.7 & 0.04 \\
\br
\end{tabular}
\item[] $^{\rm a}$ From optical to near-infrared.
\end{indented}
\end{table}

\subsection{Surveys}
We model the underlying galaxy redshift distribution with
\be
n \propto z^2 \rme^{-z/z^*},
\ee
which is in good agreement with photometry simulations based on 
degraded Hubble Deep Field North images \cite{zhan06b}.
The normalisation is adjusted to match the total surface density of 
galaxies $\bar{n}$, and the galaxy distribution peaks at $2z^*$. 
Specifications of the four surveys are listed in \tref{tab:specs}. 

Following \cite{song04}, we adopt a redshift-dependent rms shear
\be
\gamma_{\rm rms} = 0.18 + 0.042\,z
\ee
for all the surveys. Smaller rms shear may be achieved with a 
subsample of well measured galaxies, and one can optimise the trade
between $\gamma_{\rm rms}$ and $n_i$ to reduce the shape noise 
$\gamma_{\rm rms}^2 n_i^{-1}$ for WL. It can also be beneficial to 
weight galaxies differently based on their measurement 
quality such as that of the shape noise and \phz{} to optimise 
cosmological constraints.

\subsection{Dark energy constraints from BAO and WL separately}

The binning of the data affects the amount of information one can 
extract from the survey. Due to its broad kernel, WL only requires
a small number of tomographic bins to fully reach its potential. 
For DES, five bins are sufficient \cite{ma06}. We use 10 equal-width
WL bins from $z_{\rm p} = 0$ to $3.5$ for all the surveys. 
BAO, on the other hand, has narrow kernels that are defined by the
galaxy distribution $n_i$, so that one can continue to extract 
information with finer bins until shot noise overwhelms the signal
or the bin widths are less than the rms \phz{} error. 

\begin{figure}
\centering
\includegraphics[width=2.5in]{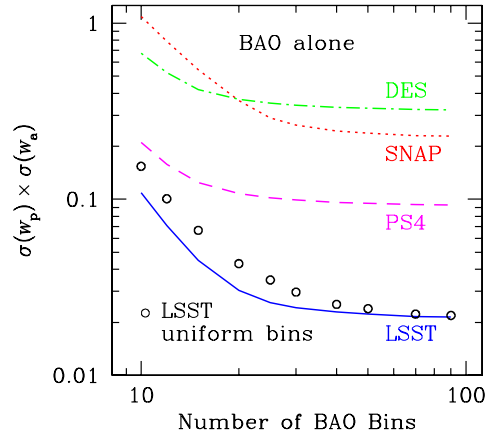}
\caption{The EP, $\sigma(w_{\rm p})\times\sigma(w_a)$, 
of BAO as a function of the number of tomographic bins for the four 
proposed surveys: DES (dash-dotted line), LSST (solid line), PS4 
(dashed line), and SNAP (dotted line). 
%rev1
The results are calculated without varying the number of 
parameters for the galaxy bias, \phz{} bias, and \phz{} rms, because 
they are defined in true-redshift space regardless the binning
of galaxies in \phz{} space.
%end
The bin widths are 
proportional to $(1+z_{\rm p})$ except that open circles are 
obtained with equal-size bins for LSST. 
The constraints are evaluated for the optimistic case in 
\tref{tab:case} but without the systematic error, which 
slightly reduces the number of bins needed to fully extract 
cosmological information from galaxy power spectra. 
\label{fig:nbin}}
\end{figure}

\Fref{fig:nbin} presents the EP as a function of the number of BAO
bins for each survey without the systematic noise, which can only 
reduce the number of bins required. We set the width of the bins to 
be proportional to $(1+z_{\rm p})$ to match the rms \phz{} error. As 
expected, one can improve the dark energy constraint by using more 
tomographic bins until it saturates. The point of saturation depends 
on the depth of the survey and $\sigma_z$. 
For LSST (solid line) and SNAP (dotted line), the EP of 40 bins is 
only $7\%$ worse than that of 90 bins, so, hereafter, we use 40 BAO 
bins for all the surveys. With uniform BAO bins (open circles), the EP
is more sensitive to the number of bins used. This suggests that it
may be useful to optimise the tomographic binning for BAO. One 
possibility is to evenly distribute the (auto power spectrum) signal 
to noise ratio for each bin.

\begin{figure}
\centering
\includegraphics[width=5in]{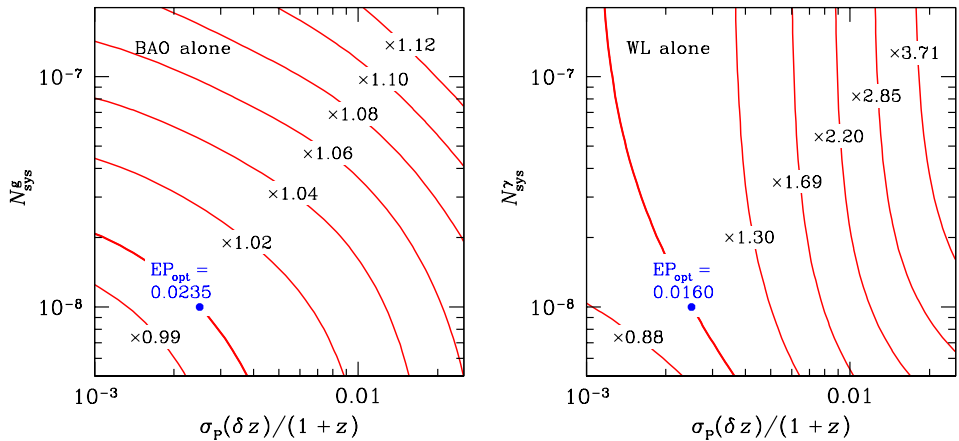}
\caption{{\it Left panel}: Contours of the EP, 
$\sigma(w_{\rm p})\times\sigma(w_a)$, of LSST BAO as a function 
of the prior of the \phz{} parameters and fiducial value 
of the simple additive systematic noise.
To reduce the dimension of the parameter space, we set the prior 
on the rms \phz{} errors to 
$\sigma_{\rm P}(\sigma_z)=\sqrt{2}\,\sigma_{\rm P}(\delta z)$.
%rev1
The solid dot locates at the coordinates specified by the 
optimistic values of $\sigma_{\rm P}(\delta z)$ 
and $N_{\rm sys}^{\rm g}$. 
%end
{\it Right panel}: Same as the left panel, but for LSST WL. Notice 
that the EP of LSST WL varies by a factor of 5.9 across the 
ranges of the \phz{} prior and systematic noise shown (from the 
lower left corner to the upper right corner), whereas 
the EP of LSST BAO changes by only $18\%$.
%rev1
If $N_{\rm sys}^{\rm g} = N_{\rm sys}^{\gamma} = 0$ and 
$\sigma_{\rm P}(\delta z)$ remains the same as that in the optimistic
case, one could achieve
${\rm EP} = 0.0231$ and $0.0068$ with BAO and WL, respectively.
%end
\label{fig:sys}}
\end{figure}

\begin{figure}
\centering
\includegraphics[width=5.1in]{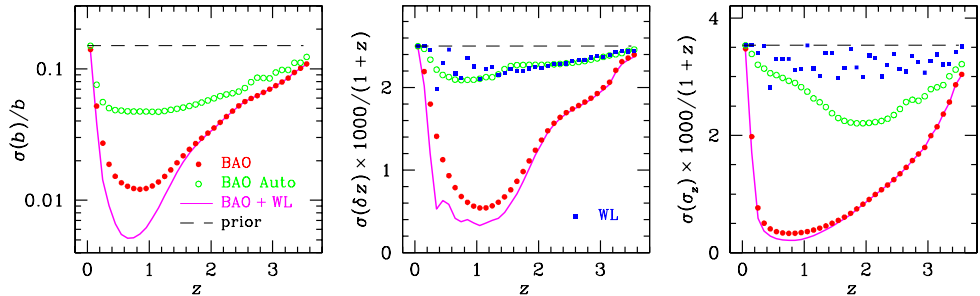}
\caption{{\it Left panel}: Constraints on the galaxy clustering bias
parameters from LSST BAO 
%rev1
(with both the auto and cross power spectra, 
solid circles), BAO with the auto power spectra only (open circles), 
%end
and joint BAO and WL (solid line). 
The $15\%$ prior on each parameter is drawn in a thin 
dashed line. The constraints are tightest near $z = 1$, roughly 
corresponding to the peak of the overall galaxy distribution.
{\it Middle panel}: Constraints on the \phz{} bias parameters.
The legends are the same as in the left panel with the 
addition of LSST WL in solid squares. The prior is taken to be
$\sigma_{\rm P}(\delta z) = 0.05 \sigma_z = 0.0025(1+z)$.
{\it Right panel}: Same as the middle panel, but for the \phz{} rms
parameters. The prior is a factor of $\sqrt{2}$ weaker than that on
$\delta z$. 
%rev1
The galaxy cross power spectra are indeed crucial for
self-calibrating the \phz{} parameters and constraining the galaxy 
clustering bias. 
%end
The oscillatory behaviour of WL constraints is 
caused by the offsets between the \phz{} parameters and each 
tomographic bin.
\label{fig:bsd}}
\end{figure}

BAO and WL, individually, are subject to different systematics and 
parameter uncertainties, but their combination can be substantially 
more robust. We first quantify in \fref{fig:sys} the effect of the
\phz{} prior and fiducial value of the additive systematic noise
on LSST BAO (left panel) and WL (right panel) using the EP. The BAO EP 
varies slowly with $\sigma_{\rm P}(\delta z)$ and 
$N_{\rm sys}^{\rm g}$, because LSST BAO is capable of self-calibrating 
the \phz{} parameters to $10^{-3}(1+z)$ 
level (see \fref{fig:bsd}) and the 
BAO systematic noise is fairly low compared to the galaxy power 
spectra. WL is unable to constrain the \phz{} parameters, so that 
it is sensitive to the \phz{} prior on $\sigma_{\rm P}(\delta z)$. 
The degradation to
WL EP caused by the shear systematic noise saturates when
$N_{\rm sys}^\gamma \gtrsim$ a few $10^{-8}$, which is consistent with 
\cite{huterer06}. 
%rev1 
\Fref{fig:sys} demonstrates that to take the advantage of the WL 
technique one needs to know the \phz{} error distribution accurately 
and have adequate control over the shear systematics.
%end

For BAO, the linear galaxy clustering bias $b$ is degenerate
with the linear growth function $G$. Therefore, one cannot extract
much useful information from the growth of the large-scale structure
with BAO. It may seem contradictory from the left panel of 
\fref{fig:bsd} that the galaxy bias $b$ (and, hence, the growth $G$) 
can actually be determined to several percent with BAO 
(solid circles). This is made 
possible by normalising the matter power spectrum to the precisely 
measured CMB fluctuations so that the difference between the galaxy 
and extrapolated matter power spectrum at low redshift gives the 
galaxy bias directly. Even though the error of the galaxy bias can
be as small as a few percent, it is still not a match to that
of the distance from BAO \cite{seo03,zhan06c}. 
%rev1
Without the galaxy cross power spectra (open circles), the BAO 
results become several times worse, showing that the cross power
spectra are useful for constraining the galaxy bias.
%end
Since WL is not affected
by the galaxy bias, its combination with BAO (solid line) improves 
$\sigma(b)$ significantly. The tightest constraint on $b$ occurs at 
a redshift that is lower than the peak of the overall galaxy 
distribution, because WL kernel peaks below the redshift of  
source galaxies. 

BAO and WL constraints on the \phz{} parameters $\delta z$ and 
$\sigma_z$ are shown, respectively, in the middle and right 
panels of \fref{fig:bsd}. One sees that WL (solid squares) 
does not have much control on the \phz{} parameters because of 
its broad kernel, while BAO, with both the auto and cross power 
spectra, is capable of calibrating the parameters to 
$10^{-3}(1+z)$ level. 
%rev1
The galaxy cross power spectra play an important role in the BAO 
self-calibration of the \phz{} error distribution; without them, 
the BAO constraints on the \phz{} parameters become much weaker.
%end
\Fref{fig:bsd} thus explains the 
markedly different response to the \phz{} bias prior between the BAO 
and WL EPs in \fref{fig:sys}. It should be emphasised that 
the BAO self-calibration of the \phz{} parameters cannot replace 
spectroscopic calibrations, because, without knowing how to 
faithfully parametrize the \phz{} error distribution, 
the self-calibration will be less useful. 

The \phz{} constraints are much tighter 
than those of the galaxy clustering bias. The reasons are that the 
galaxy power spectra are more sensitive to the \phz{} parameters than 
to the galaxy bias and that the dependence of the galaxy power spectra
on the \phz{} parameters is very unique. For instance, the derivative 
$\rmd \ln \bar{n}_i\bar{n}_j P_{ij}^{\rm gg} / \rmd b$ is
roughly unity at $z \sim 1$ for all $i$ and $j$, whereas 
$\rmd \ln \bar{n}_i\bar{n}_j P_{ij}^{\rm gg} / \rmd \delta z$ varies
from $-4$ to $4$ and behaves distinctively different from one pair 
of $i$ and $j$ to another (\fref{fig:cldcl}).

\subsection{Joint analysis of BAO and WL}

\begin{figure}
\centering
\includegraphics[width=2.5in]{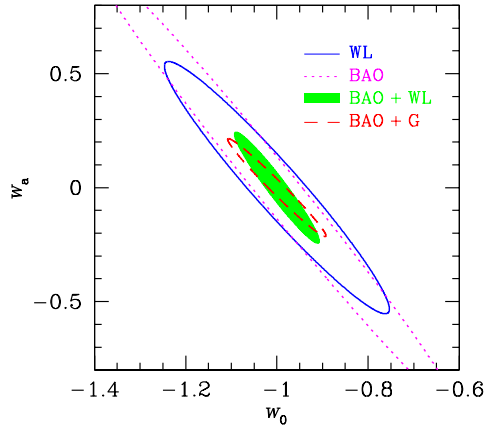}
\caption{Complementarity between BAO and WL. The $1\sigma$ 
$w_0$--$w_a$ contours are shown for LSST BAO (dotted line), 
LSST WL (solid line), joint BAO and WL (shaded area), and the case of 
BAO plus growth information (dashed line), for which we fix the galaxy 
clustering bias parameters to their fiducial values. The last case is
unrealistic, but it demonstrates the importance of the 
growth information.
\label{fig:w0wa}}
\end{figure}

\begin{figure}
\centering
\includegraphics[width=5in]{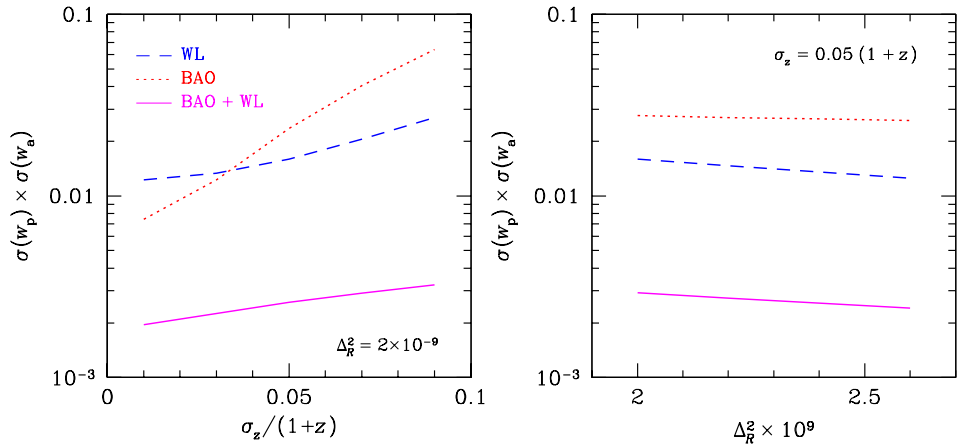}
\caption{{\it Left panel}: The EP as a function of the rms \phz{} 
error $\sigma_z$ for LSST BAO (dotted line), LSST WL (dashed line), 
and the two combined (solid line). We have applied the 
same tomographic binning regardless the value of $\sigma_z$. 
With finer bins, BAO results can be further improved at small 
$\sigma_z$, although WL and the combined results will remain less 
sensitive to $\sigma_z$.
{\it Right panel}: The EP as a function of the 
normalisation of the primordial curvature fluctuation $\Delta_R^2$
at $k = 0.05\,\mbox{Mpc}^{-1}$.
As $\Delta_R^2$ increases, the galaxy number density fluctuation and
weak lensing shear signal increase. If the signal is dominant over 
the noise (in a single mode), then the constraints on $w_0$ and 
$w_a$ will not change with the normalisation, 
whereas in the opposite case $\sigma(w_0)$ and $\sigma(w_a)$ will 
be inversely proportional to the normalisation. To 
isolate the effect of the normalisation, we use the same maximum 
multipole of BAO, which is determined for 
$\Delta_R^2 = 2.6\times 10^{-9}$, for all normalisation cases. 
This means that the constraints with $\Delta_R^2 < 2.6\times 10^{-9}$
are calculated from fewer multipoles than what would otherwise be 
included with the criteria $\Delta_\delta^2(k;z) < 0.4$ and 
$40 \le \ell \le 3000$. 
\label{fig:sigz}}
\end{figure}

Based on \fref{fig:sys} and \fref{fig:bsd}, one expects a great 
improvement of errors on cosmological parameters when BAO and WL are 
analysed jointly, which is indeed demonstrated in 
\fref{fig:w0wa}. The 
combination (shaded area) reduces the EP by more than a factor of 6 
compared to that of LSST WL alone (solid line). This is more than
the factor of $\sim \sqrt{2}$ reported in \cite{hu04b}, which does
not include systematic errors or uncertainties in the \phz{} error
distribution. LSST BAO (dotted line) does not
seem competitive to WL at all, if assessed by $\sigma(w_0)$ and 
$\sigma(w_a)$ only. Since the errors in $\sigma(w_0)$ and 
$\sigma(w_a)$ are highly correlated, the EP of BAO is merely
a factor of $1.47$ larger than that of WL (see \fref{fig:sigz} and 
\tref{tab:opt}).
The important issue here is not that which technique is better, 
but rather that one must analyse BAO and WL data (and possibly 
others) {\it jointly} to overcome the weakness of each.

To understand the role of the growth function, we artificially fix
the galaxy clustering bias parameters for BAO, which enables BAO to 
utilise the growth information. Since the growth function and galaxy 
bias are degenerate, the error of the growth function, if 
similarly parametrized, is the same as that of the galaxy bias. 
However, when the galaxy bias is fixed, the error of the growth 
function will be much smaller than that of the galaxy bias in 
\fref{fig:bsd}. The resulting constraints (dashed 
line in \fref{fig:w0wa}) on $w_0$ and $w_a$ are, not surprisingly, 
as tight as those of BAO and WL combined, because the narrow kernel 
of BAO makes it possible to sample the cosmic density field at very 
fine intervals in redshift \cite[see also][]{knox06c}. We note, 
however, that the determination of the (product of the 
galaxy clustering bias and) growth and distance are correlated. 
If the distance is poorly measured, the growth is not likely to 
be known very well either. Conversely, an accurate determination 
of the growth can improve the distance measurements, because
the amplitude of the BAO and WL power spectra depend on the 
distance (not just distance ratios).

Error forecasts inevitably depend on the assumed fiducial model.
For example, because of its narrow kernel, BAO is sensitive to the 
rms \phz{} error $\sigma_z$, which can vary greatly with the galaxy 
population, filter design, survey depth, and so on.
Another parameter of interest is the normalisation of the matter
power spectrum, which is proportional to the normalisation 
$\Delta_R^2$ of the primordial curvature power spectrum at 
$k = 0.05\,{\rm Mpc}^{-1}$. Accompanying
a lower estimate of the electron optical depth to the last 
scattering surface $\tau = 0.09$ and a slight tilt 
$n_{\rm s} = 0.951$, the recent analysis of the three-year 
{\it WMAP} date lowers the value of $\Delta_R^2$ by $25\%$.
This reduces the amplitude of BAO and WL power spectra and can result
in a weakening of the parameter constraints by up to $25\%$ if the 
shot (shape) noise and systematic noise are dominant over the 
(auto power spectra) signal in a single mode. 

\Fref{fig:sigz} quantifies the dependence of the EP on $\sigma_z$ 
(left panel) and $\Delta_R^2$, (right panel). The sensitivity 
of BAO (dotted line) and, to a lesser degree, WL (dashed line) EPs to 
$\sigma_z$ is clearly seen, and the combination (solid line) is
less sensitive to $\sigma_z$. From the right panel, one sees that
the BAO EP does not change much with the normalisation 
$\Delta_R^2$. This means that it is not limited by noise, but rather 
by the \phz{} error: with 40 bins whose widths are proportional to 
$(1+z_{\rm p})$, the bins are already narrower than the \phz{} rms.
The WL EP  improves by $28\%$, i.e., roughly $14\%$ each
on $w_{\rm p}$ and $w_a$, as the normalisation increases from 
$\Delta_R^2 = 2\times10^{-9}$ to $ 2.6\times10^{-9}$, indicating 
that the noise is more important for WL but not yet completely 
dominant.

\begin{figure}
\centering
\includegraphics[width=5in]{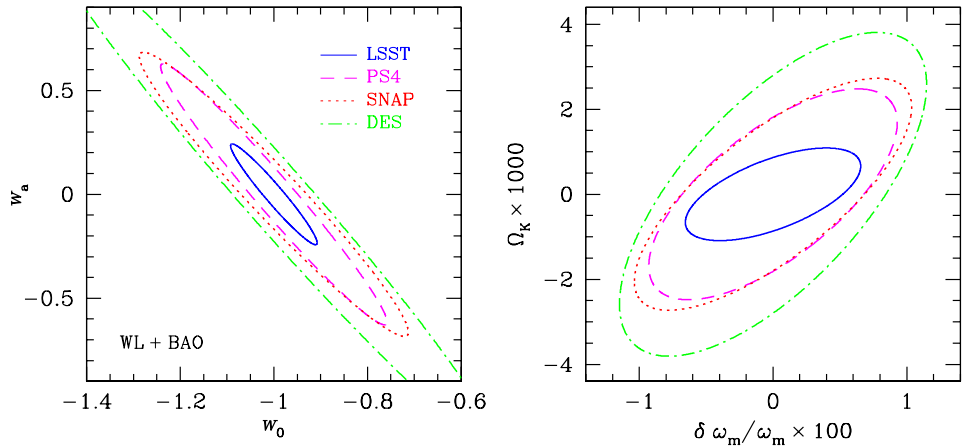}
\caption{{\it Left panel}: Joint BAO and WL constraints on the dark 
energy EOS parameters $w_0$ and $w_a$ for DES (dash-dotted line), 
LSST (solid line), PS4 (dashed line), and SNAP (dotted line). 
{\it Right Panel}: Same as the left panel, but for the matter density
$\omega_{\rm m}$ and curvature term $\Omega_{\rm K}$. Note that 
{\it Planck} alone constrains $\omega_{\rm m}$ to $\sim 1\%$ (see 
\tref{tab:opt}) and that SN and cluster counting can further 
improve the constraints.
\label{fig:lspd}}
\end{figure}

\subsection{Forecasts}
We present $w_0$--$w_a$ and $\omega_{\rm m}$--$\Omega_{\rm K}$ 
$1\sigma$ error
contours of joint BAO and WL analysis for the four surveys in 
\fref{fig:lspd} 
and list the $1\,\sigma$ error bounds on all the cosmological 
parameters in \tref{tab:opt} (optimistic case) and 
\tref{tab:pes} (pessimistic case).

\Fref{fig:lspd} demonstrates that a deep and wide survey like LSST 
can tighten the constraints on dark energy and other cosmological 
parameters substantially. 
In its wide and shallow survey mode, PS4 can perform 
equally well as SNAP. Of course, the SNAP SN program will provide 
additional
constraints that are fairly complementary to BAO and WL  
\cite[e.g.,][]{knox06c}. Likewise, the three ground-based surveys 
all plan to carry out SN and/or cluster counting to some degree. 

With $w = -1$, the combination of {\it WMAP} and the Supernova 
Legacy Survey data determines $\Omega_{\rm K}$ at a few percent level
\cite{spergel06}. The right panel of \fref{fig:lspd} shows that 
LSST BAO+WL and {\it Planck} can further reduce the uncertainty 
to $10^{-3}$ with our parametrization of  the dark energy EOS, 
which may lead to interesting tests of 
inflation \cite{knox06a,knox06c}. It is also possible to measure the
mean curvature from BAO and WL without assuming the dynamics and 
content of the universe \cite{bernstein06}. Though less constraining,
such a model independent method can provide a very useful check. 

\fulltable{\label{tab:opt}\noindent Error forecasts 
with optimistic BAO and WL survey parameters}
\lineup  \small
\begin{tabular}{@{}l*{2}{@{\hspace{1ex}}c}*{3}{@{\hspace{1em}}c}
@{\hspace{0.4ex}}*{2}{@{\hspace{1ex}}c}@{\hspace{-0.3ex}}
*{6}{@{\hspace{1em}}c}@{\hspace{0.5ex}}c@{}}
\br 
 & & & & & & $\ln \omega_{\rm m}$ & $\ln \omega_{\rm b}$ & 
$\theta_{\rm s}$ & $\Omega_{\rm K}$ & $\tau^{\,\rm c}$ & 
$Y_{\rm p}$ & $n_{\rm s}$ & $\alpha^{\,\rm d}$ & 
$\ln \Delta_R^{2\,\rm c}$ \\ \ns  & & & & & & \crule{9} \\
Survey & {\sc bao} & {\sc wl} & $w_{\rm p}^{\,\rm a}$ & 
$w_0^{\,\rm b}$ & $w_a^{\,\rm b}$ & \centre{9}{($10^{-3}$)} \\
\mr
{\it Planck} & & & 0.47 & 1.1 & 2.5 & 9.5 & 11 & 
0.19 & 17\0 & 9.4 & 16 & 9.8 & 8.1\0 & 19 \\
%0.465377 1.06144 2.46997 0.00954584 0.0105299 0.000187054 0.0170052 0.00938863 0.01594 0.00982984 0.00813677 0.0188733
% sigma[ln(wm)] & sigma[ln(wb)] already
\mr
 & $\times$ & & 0.037 & 0.26\0 & 0.64 & 5.1 & 6.5 & 0.14 & 1.1\0 & 
8.9 & 7.0 & 4.2 & 0.81 & 18 \\
%0.023476 0.260214 0.642586 0.000644465 0.000145047 0.000137229 0.00111158 0.00887438 0.00695627 0.00419145 0.000809585 0.0176763
% sigma(wm) & sigma(wb)
LSST & & $\times$ & 0.044 & 0.16\0 & 0.36 & 8.5 & 8.9 & 0.16 & 
2.5\0 & 8.8 & 10 & 6.5 & 0.73 & 18 \\
%0.0159821 0.162618 0.364982 0.00108343 0.000197448 0.000156702 0.00253754 0.00877212 0.0101473 0.00649673 0.00072776 0.0176802
 & $\times^{\rm e}$ & $\times$ & 0.016 & 0.061 & 0.16 & 4.3 & 5.8 & 
0.13 & 0.72 & 6.1 & 5.4 & 2.7 & 0.41 & 12\\
%0.00260153 0.0611039 0.160141 0.000549045 0.000129832 0.000130315 0.000718161 0.00605798 0.00542497 0.00270997 0.000414271 0.0115031
\mr
 & $\times$ & & 0.075 & 0.49\0 & 1.3\0 & 6.8 & 8.5 & 0.16 & 3.1\0 & 
9.0 & 11 & 6.8 & 2.1 & 18 \\
%0.0971711 0.492072 1.30211 0.000865904 0.000188606 0.000157276 0.00308937 0.00897323 0.0107147 0.00681055 0.00210933 0.0179065
PS4 & & $\times$ & 0.083 & 0.52\0 & 1.2\0 & 8.8 & 9.2 & 0.16 & 
3.3\0 & 9.1 & 11 & 7.2 & 1.8 & 18 \\
%0.103213 0.519666 1.23719 0.00111349 0.000205307 0.000162005 0.00325278 0.00913895 0.0113289 0.00722521 0.00180174 0.0184002
 & $\times$ & $\times$ & 0.034 & 0.16\0 & 0.41 & 6.1 & 7.4 & 0.14 & 
1.6\0 & 8.1 & 8.3 & 5.0 & 1.3 & 16 \\
%0.0139511 0.159487 0.414079 0.000778145 0.000165224 0.000144229 0.00163477 0.00810986 0.00828556 0.00497316 0.00129548 0.016067
\mr
 & $\times$ & & 0.12\0 & 0.84\0 & 2.1\0 & 7.7 & 8.8 & 0.16 & 3.4\0 & 
9.0 & 11 & 7.1 & 2.2 & 18 \\
%0.247982 0.844856 2.06026 0.000983983 0.000197039 0.0001606 0.00335891 0.00898889 0.0109978 0.00706359 0.0022442 0.017899
SNAP & & $\times$ & 0.11\0 & 0.44\0 & 0.86 & 8.8 & 9.2 & 0.16 & 
2.9\0 & 9.1 & 11 & 7.2 & 1.8 & 18 \\
%0.0989123 0.437052 0.863234 0.00111182 0.000204779 0.000161678 0.00292629 0.00910439 0.0112552 0.00718155 0.0017884 0.0183188
 & $\times$ & $\times$ & 0.046 & 0.19\0 & 0.45 & 6.8 & 7.6 & 0.15 & 
1.8\0 & 8.4 & 8.2 & 5.0 & 1.2 & 16 \\
%0.0209547 0.189541 0.450812 0.00086837 0.000169526 0.000145208 0.00179906 0.00838249 0.0082344 0.00501484 0.00123302 0.0164667
\mr
 & $\times$ & & 0.14\0 & 0.90\0 & 2.4\0 & 8.0 & 9.1 & 0.17 & 5.5\0 & 
9.0 & 12 & 7.8 & 3.6 & 18 \\
%0.331252 0.900156 2.36612 0.00101115 0.000203924 0.000166271 & 0.00551926 & 0.0090238 & 0.0124734 & 0.00780132 & 0.00355379 & 0.018\\
DES & & $\times$ & 0.16\0 & 1.0\0\0 & 2.4\0 & 8.8 & 9.3 & 0.17 & 
4.6\0 & 9.2 & 12 & 7.6 & 3.2 & 18 \\
%0.37626 & 1.00268 & 2.38013 & 0.00112187 & 0.000207629 & 0.000165428 & 0.00463267 & 0.00918005 & 0.0120763 & 0.00762295 & 0.00319624 & 0.018477\\
 & $\times$ & $\times$ & 0.058 & 0.31\0 & 0.80 & 7.6 & 8.4 & 0.15 & 
2.5\0 & 8.7 & 10 & 6.4 & 2.3 & 17 \\
%0.0466353 0.312057 0.797495 0.000961494 & 0.00018782 & 0.000154993 & 0.00251092 & 0.00867921 & 0.0103748 & 0.00640668 & 0.00234469 & 0.0173105\\
\br
\end{tabular} \\

\parbox{\textwidth}{ \small
\noindent $^{\rm a}$ The error of the dark energy EOS at the pivot 
epoch [equation \eref{eq:ep}], $\sigma(w_{\rm p})$, 
is uncorrelated with that of $w_a$.

\noindent $^{\rm b}$ When combining CMB with BAO and WL, we do not 
include the CMB constraints on $w_0$ and $w_a$,
which are sensitive to the late-time ISW effect on large scales
\cite{hu02a}.  % 4.12238        14.4437  ell $>$ 40 \\

\noindent $^{\rm c}$ If the polarisation data at $2 \le \ell < 10$
are included, {\it Planck} alone can achieve $\sigma(\tau) = 0.0049$
and $\sigma(\ln \Delta_R^2) = 0.010$.

\noindent $^{\rm d}$ The running 
$\alpha = \rmd \ln n_{\rm s}(k) / \rmd \ln k$ at 
$k_{\rm f} = 0.05\,{\rm Mpc}^{-1}$, and the primordial curvature 
power spectrum $\Delta_R^2(k) = \Delta_R^2(k_{\rm f}) 
(k / k_{\rm f})^{n_{\rm s}(k_{\rm f}) - 1 + 
0.5 \alpha \ln (k / k_{\rm f})}$ is accurate to second order.

\noindent $^{\rm e}$ Without the self-calibration of \phz{} 
parameters from the galaxy cross power spectra, 
the dark energy constraints would degrade to
$\sigma(w_{\rm p}) = 0.021$, $\sigma(w_0) = 0.094$, and 
$\sigma(w_a) = 0.22$. 
%rev1
One could achieve $\sigma(w_{\rm p}) = 0.012$, 
$\sigma(w_0) = 0.048$, and $\sigma(w_a) = 0.13$, if
$N_{\rm sys}^{\rm g} = N_{\rm sys}^{\gamma} = 0$. 
%end
}
\endfulltable

\fulltable{\label{tab:pes}\noindent Error forecasts 
with pessimistic BAO and WL survey parameters}
\lineup  \small
\begin{tabular}{@{}l*{2}{@{\hspace{1ex}}c}*{3}{@{\hspace{1em}}c}
@{\hspace{0.4ex}}*{2}{@{\hspace{1ex}}c}@{\hspace{-0.3ex}}
*{6}{@{\hspace{1em}}c}@{\hspace{0.5ex}}c@{}}
\br 
 & & & & & & $\ln \omega_{\rm m}$ & $\ln \omega_{\rm b}$ & 
$\theta_{\rm s}$ & $\Omega_{\rm K}$ & $\tau$ & $Y_{\rm p}$ & 
$n_{\rm s}$ & $\alpha$ & $\ln \Delta_R^2$ \\ \ns
 & & & & & & \crule{9} \\
Survey & {\sc bao} & {\sc wl} & $w_{\rm p}$ & $w_0$ & $w_a$ & 
\centre{9}{($10^{-3}$)} \\
\mr
 & $\times$ & & 0.039 & 0.28\0 & 0.69 & 5.2 & 6.7 & 0.14 & 1.2\0 & 
9.0 & 7.3 & 4.4 & 0.90 & 18 \\
%0.0266212 & 0.278282 & 0.689445 & 0.000662033 & 0.000148963 & 0.000138967 & 0.00122867 & 0.00896538 & 0.00727456 & 0.00444831 & 0.000898265 & 0.01787 \\
LSST & & $\times$ & 0.067 & 0.28\0 & 0.57 & 8.6 & 8.9 & 0.16 & 
2.6\0 & 8.9 & 10 & 6.6 & 0.74 & 18 \\
%0.0386529 & 0.282987 & 0.572931 & 0.0010882 & 0.000198423 & 0.000157203 & 0.00257801 & 0.00888702 & 0.010234 & 0.00656619 & 0.000740535 & 0.0179\\
 & $\times$ & $\times$ & 0.019 & 0.069 & 0.18 & 4.4 & 5.9 & 0.13 & 
0.78 & 6.3 & 5.6 & 2.9 & 0.44 & 12 \\
%0.00338532 & 0.0692371 & 0.176621 & 0.000564769 & 0.000131827 & 0.000131202 & 0.000779223 & 0.00625028 & 0.00555377 & 0.00285032 & 0.000441175 & 0.0119\\
\mr
 & $\times$ & & 0.078 & 0.52\0 & 1.4\0 & 7.0 & 8.6 & 0.16 & 3.3\0 & 
9.1 & 11 & 7.0 & 2.5 & 18 \\
%0.107921 & 0.518544 & 1.37536 & 0.000882665 & 0.000191196 & 0.000159025 & 0.00329067 & 0.00906321 & 0.0110953 & 0.00702138 & 0.00250558 & 0.018\\
PS4 & & $\times$ & 0.13\0 & 0.90\0 & 2.0\0 & 8.8 & 9.2 & 0.16 & 
3.7\0 & 9.1 & 11 & 7.3 & 2.0 & 18 \\
%0.269985 & 0.900293 & 2.00171 & 0.00111457 & 0.000205589 & 0.000162347 & 0.0037324 & 0.00914839 & 0.0114003 & 0.00726599 & 0.00195907 & 0.01841\\
 & $\times$ & $\times$ & 0.037 & 0.18\0 & 0.46 & 6.3 & 7.6 & 0.15 & 
1.8\0 & 8.2 & 8.6 & 5.2 & 1.5 & 16 \\
%0.017139 & 0.178946 & 0.463679 & 0.000801698 & 0.00016904 & 0.000145917 & 0.00176172 & 0.00824519 & 0.00861191 & 0.00521665 & 0.00146544 & 0.0163898\\
\mr
 & $\times$ & & 0.13\0 & 0.92\0 & 2.2\0 & 7.8 & 8.9 & 0.16 & 3.6\0 & 
9.1 & 1.1 & 7.2 & 2.4 & 18 \\
%0.288717 & 0.91822 & 2.23856 & 0.000993897 & 0.000198671 & 0.000161663 & 0.00360614 & 0.00906365 & 0.0112161 & 0.00719126 & 0.00237271 & 0.018\\
SNAP & & $\times$ & 0.14\0 & 0.53\0 & 1.0\0 & 8.8 & 9.2 & 0.16 & 
3.0\0 & 9.1 & 11 & 7.2 & 1.8 & 18 \\
%0.138185 & 0.52666 & 1.01785 & 0.00111244 & 0.000204866 & 0.000161754 & 0.00298149 & 0.00911293 & 0.0112688 & 0.00719019 & 0.00182668 & 0.01833\\
 & $\times$ & $\times$ & 0.057 & 0.23\0 & 0.52 & 7.0 & 7.7 & 0.15 & 
1.9\0 & 8.5 & 8.4 & 5.2 & 1.3 & 17 \\
%0.0299382 & 0.225562 & 0.522646 & 0.00088999 & 0.000172242 & 0.000146441 & 0.00190819 & 0.0084881 & 0.00843331 & 0.00517617 & 0.00131167 & 0.0167414\\
\mr
 & $\times$ & & 0.15\0 & 0.99\0 & 2.6\0 & 8.1 & 9.2 & 0.17 & 6.1\0 & 
9.1 & 13 & 8.0 & 4.1 & 18 \\
%0.395753 & 0.99204 & 2.61321 & 0.00102662 & 0.000205596 & 0.000168256 & 0.00606463 & 0.00912455 & 0.0128977 & 0.00802101 & 0.00413067 & 0.018\\
DES & & $\times$ & 0.22\0 & 1.4\0\0 & 3.2\0 & 8.8 & 9.3 & 0.17 & 
5.3\0 & 9.2 & 12 & 7.7 & 3.4 & 18 \\
%0.70202 & 1.43606 & 3.2066 & 0.00112374 & 0.000208088 & 0.000166153 & 0.00532458 & 0.00918665 & 0.0122253 & 0.0077034 & 0.00343071 & 0.018\\
 & $\times$ & $\times$ & 0.065 & 0.35\0 & 0.90 & 7.7 & 8.6 & 0.16 & 
2.7\0 & 8.8 & 11 & 6.6 & 2.7 & 18 \\
%0.0585 & 0.353276 & 0.903767 & 0.000981849 & 0.000190835 & 0.000156769 & 0.00271857 & 0.0087978 & 0.0107073 & 0.00662804 & 0.00265079 & 0.017603\\
\br
\end{tabular}
\endfulltable

Our forecasts for {\it Planck} are given in \tref{tab:opt} for 
reference. CMB is affected by dark energy mostly through
the late-time ISW effect on large angular scales \cite{hu02a}. If we 
limit the multipoles to $\ell \ge 40$ for {\it Planck}, then its EP 
increases to $11$. Since the properties of the cosmic density field 
on very large scales have not been measured very well yet,
we neglect {\it Planck} constraints on the dark energy EOS parameters
when combining it with BAO and WL. This treatment does not affect the 
combined results, especially those of all the three techniques, 
because $w_0$ and $w_a$ are better measured by BAO and WL.
For non-dark-energy parameters (excluding $\Omega_{\rm K}$ and 
$\alpha$), the constraints are not so easily improved by BAO and WL
except for LSST, which shows again that the 
deep photometry and wide area are not merely a technical challenge 
but rather a requirement for BAO and WL to be competitive with future
CMB measurements beyond the dark energy EOS \cite{zhan06b}.

As a result of its fine sampling of the density field in redshift 
space, BAO does relatively well in measuring parameters 
($\omega_{\rm m}$ and $\omega_{\rm b}$) that affect the features 
in the power spectrum. Since noises are more dominant for 
WL (see \fref{fig:sys} and \fref{fig:sigz}), 
BAO can perform better than WL for
shallow surveys, especially in the pessimistic case 
(\tref{tab:pes}). BAO has another slight advantage 
that it probes the universe at a little higher redshift than WL 
does, because the lensing kernel peaks below the redshift of  
source galaxies. This advantage fades away when the survey is 
shallow and/or narrow, and it is a part of the reason that the 
relative strength of the BAO
and WL constraints on $\Omega_{\rm K}$ inverts from LSST to DES.
For SNAP, its high galaxy number density (low shape noise) allows
WL to remain significantly more competitive than BAO in the 
pessimistic case.

Our errors of the running $\alpha$ for LSST WL and SNAP WL are 
considerably tighter than those in \cite{song04}, but
the errors for DES BAO are larger than those forecasted 
for a $f_{\rm sky} = 0.1$ and $z < 2.3$ BAO survey in \cite{dolney06}.
Since we do not include neutrino mass as \cite{song04} does, and 
since \cite{dolney06} assumes a survey with many more galaxies 
than DES, our results can still be consistent with theirs. Despite the 
differences, these forecasts all predict fairly tight constraints on
$\alpha$, which is expected to be of the same order as
$(n_{\rm s} - 1)^2$ from simple inflation models. 
With $n_{\rm s} - 1 = 0.05$ \cite{spergel06} and 
measurements of $\alpha$ at $10^{-3}$ level from future BAO and WL 
surveys, we will have another interesting test for inflation 
\cite{song04}.

Comparing the optimistic forecasts with pessimistic ones, we find 
that the joint BAO and WL results are fairly robust to the priors on 
non-cosmological parameters and to simple additive systematics. This
clearly demonstrates the necessity of utilising multiple techniques
jointly to extract cosmology information. 

\section{Discussion and Conclusions} \label{sec:dis}

We have shown that BAO and WL are highly complementary to each other
especially in the presence of the uncertainties of the \phz{} error 
distribution and inaccurate knowledge of the galaxy clustering bias.
For non-Gaussian \phz{} errors, more parameters will be needed to 
characterise the error distribution, which inevitably leads to 
weaker constraints on these parameters \cite{schneider06}. 
The impact of non-Gaussian \phz{} errors on the 
forecasts of cosmological parameters will be investigated separately. 

We emphasise that one cannot rely solely on the galaxy power 
spectra for \phz{} information. 
One may notice in \fref{fig:bsd} that the galaxy clustering bias 
and \phz{} parameters are not constrained by the galaxy power 
spectra at both the low-redshift end, where very few power spectrum
modes can be used, and the high-redshift end, where the data are poor. 
With long-tail \phz{} error models, the unconstrained errors at
the two redshift ends can easily propagate to the whole redshift 
range and diminish the BAO ability to self-calibrate the \phz{} 
parameters. In order to achieve 
the BAO self-calibration of \phz{} errors, one must know 
how to parametrize the probability distribution of 
\phz{} errors faithfully. 
In other words, it is crucial to map the \phz{} error 
distribution and reduce its uncertainties as much as possible
with massive spectroscopic calibrations 
{\it and} state-of-the-art \phz{} algorithms.

Even without knowing the true \phz{} error distribution 
accurately, one can still
use the tomographic BAO and WL power spectra to check for deviations
from the assumed \phz{} error distribution and to detect certain
systematic errors. For example, a detection of the cross power 
spectrum between a very low redshift BAO bin and a high redshift one 
above the expected level could be a result of catastrophic \phz{} 
errors between the bins or residuals of extinction correction. 
The effect of the former would be confined between the bins around
the two redshifts, while that of the latter would be likely to 
spread over all bins. In any case, the cross power spectra arising
from the two causes would have distinct shapes.
Similarly, an unexpected detection of the cross power spectrum 
between a low redshift WL bin and a higher redshift BAO bin could be 
due to some systematics.

%rev2 
% 1 paragraph moved from section 4.2 with modification
We have assumed 
that the galaxy bias is not correlated with \phz{} errors. 
In reality, different types of galaxies have different spectral
features that affect their \phz{} estimation and, meanwhile, do
not cluster with the same strength. Therefore,
\phz{} errors are inevitably coupled with the galaxy bias 
through the galaxy type. More realistic forecasts should include 
multiple types of galaxies and assign separate galaxy bias and 
\phz{} parameters for each type. Since the shot noise is relatively 
low for BAO, it is practical to split the galaxies into 
a few groups for better analyses. 
One may actually improve the \phz{} constraints with multiple galaxy
types \cite{schneider06}.
%end

One may worry about the break-down of the Limber approximation on 
large angular scales. The break-down
of an approximation is not a loss of information, but, rather, a 
demand for proper means to extract the information.
For parameter estimation, full calculations should replace
the Limber approximation on large angular scales, but for error
estimation, we can make more conservative cuts in $\ell$ to see the
effect of completely ignoring the large-scale information. 
If we set the minimum multipole to $\ell_{\rm min} = 100$, which
removes the information of the broadband shape of the matter power
spectrum (see \fref{fig:cldcl}), the constraints on $w_0$ and 
$w_a$ degrade by less than $20\%$ for LSST BAO and less than $13\%$
for the joint analysis. In this test, BAO bins below 
$z_{\rm p} = 0.23$ have no multipole usable. 
BAO errors on $w_0$ and $w_a$ will double if one also excludes the 
first few acoustic peaks by setting $\ell_{\rm min} = 300$.
In contrast, WL errors will only 
increase by $25\%$ with $\ell_{\rm min} = 300$. 
These tests confirm that the constraining power of 
the broadband shape of the matter power spectrum is sub-dominant to 
that of the acoustic oscillations for the BAO technique \cite{seo03}.

On the other end of the scales, one may worry about our somewhat
aggressive maximum multipoles $\ell_{\rm max} = 3000$ for BAO and
$\ell_{\rm max} = 2000$ for WL.
We find that a cut of $\ell_{\rm max} = 2000$ for LSST BAO increases 
the error on $w_0$ and $w_a$ by less than $2\%$. This is not 
surprising, because the angular galaxy power spectra of the 15 
($z_{\rm p} > 1.7$) BAO bins that are affected by this cut are 
practically featureless at $2000 < \ell \le 3000$, i.e., 
these multipoles do not contribute much to distance measurements. 
With $\ell_{\rm max} = 2000$ for BAO and $\ell_{\rm max} = 1000$ 
for WL, the LSST BAO and WL joint constraints on $w_0$ and $w_a$ 
weaken by only $8\%$. Hence,
we conclude that, given the survey parameters and assumptions about
systematics, our forecasts are fairly robust against the range of
multipoles included in the analysis.

\begin{figure}
\centering
\includegraphics[width=2.5in]{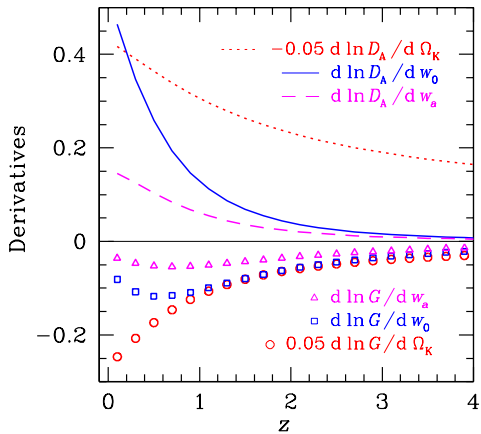}
\caption{Derivatives of the comoving angular diameter distance 
$D_{\rm A}$ and linear growth function $G$ with respect to $w_0$,
$w_a$, and $\Omega_{\rm K}$. When taking the derivatives, we hold 
the angular size of the sound horizon at the last scattering surface
$\theta_{\rm s}$ fixed. At $z \sim 0$, the distance derivatives are 
effectively derivatives of the implicit Hubble constant with respect 
to the parameters. The linear growth function is normalised such that 
$G(a) = a$ in an Einstein-de Sitter universe [as opposed to G(1) = 1
regardless the cosmological model].
\label{fig:dddg}}
\end{figure}

There have been interesting discussions about dark energy 
constraints from measurements of the distance and the growth of 
the large-scale structure.
%rev1
The results, depending on how the distance and growth are separated, 
range from that they are equally powerful \cite{simpson05,zhang05} 
to that the constraints from the growth function are 
much weaker than those from the distance \cite{abazajian03,knox06b}.
The relative strength of the distance and growth function in 
constraining dark energy depends on how sensitive 
the distance and growth function are to the parameters and how
well they can be reconstructed from WL data.
%end

We plot the distance and growth derivatives with respect to 
$w_0$, $w_a$, and $\Omega_{\rm K}$ in 
\fref{fig:dddg}. Since the growth function 
(open symbols) is less sensitive 
to the parameters than the distance (lines) and it is not so well 
determined as the distance with WL \cite{knox06b}, dark energy 
constraints from the growth are sub-dominant to those from the 
distance.
%rev1
However, this should not undermine the advantage of being able
to measure the growth function (or, more appropriately, the 
amplitude of the matter power spectrum).
%end
To measure the distance with WL, one needs to know how the 
power spectrum amplitude changes with redshift, because 
the shear power spectrum depends on both the lensing geometry 
and growth function. In fact, the WL EP will 
increase by more than 10 times, if one gives up the ability to
measure the growth function by applying artificial bias 
parameters [with the fiducial 
value $b = 1$ and $\sigma_{\rm P}(\ln b) = 0.2$] to the shear
power spectra \cite{knox06c}. 
Conversely, it is not possible to measure the growth function
accurately if the distance is not, because one needs to know the 
scale $k = \ell / D_{\rm A}$ accurately to estimate the amplitude
of the power spectrum.

The galaxy power spectra, being able to use the 
BAO features in the matter power spectrum, can better isolate the 
distance from (the product of the galaxy bias and) growth 
\cite{zhan06c}, which partially explains the $\lesssim 10\%$ 
difference in the BAO constraints between the optimistic case 
[\tref{tab:opt} with $\sigma_{\rm P}(\ln b) = 0.15$] and 
pessimistic case [\tref{tab:pes} with $\sigma_{\rm P}(\ln b) = 0.3$].
%rev1
On the other hand, if the galaxy bias was known perfectly,
then one could determine the distance
with both the amplitude and BAOs of the galaxy power spectra. 
This reduces errors in both the distance and growth function and 
leads to the tight dark energy constraints of the BAO+G case in 
\fref{fig:w0wa}.
%end

\begin{figure}
\centering
\includegraphics[width=2.5in]{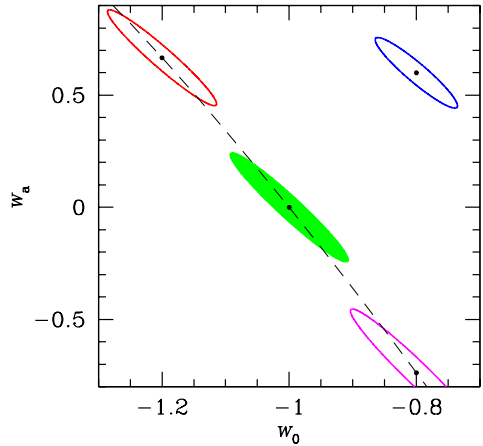}
\caption{Error contours in the $w_0$--$w_a$ plane with joint 
LSST BAO and WL analysis. 
Solid dots mark the fiducial values of $w_0$ and $w_a$. 
All the four models have the same fiducial values of the other 9 
cosmological parameters . 
The effects of $w_0$ and $w_a$ on the angular size of
the sound horizon at the last scattering surface $\theta_{\rm s}$
cancel each other along the thin dashed line. In other words, we
require that the two implicit parameters $\Omega_{\rm X}$ and $h$ 
be the same for models on this line. The supergravity-inspired
model at $(w_0,w_a) = (-0.8, 0.6)$ 
has $\Omega_{\rm X} = 0.61$ and $h=0.57$.
\label{fig:eos}}
\end{figure}

\Fref{fig:dddg} also illustrates why SN constraints on dark energy 
are sensitive to the flatness  assumption \cite{linder05b,knox06c}. 
The reason is that distance is much more sensitive to 
curvature then to the dark energy EOS. In addition, the (negative)
distance 
derivative with respective to $\Omega_{\rm K}$ (dotted line) has a 
redshift dependence that somewhat resembles 
$\rmd \ln D_{\rm A}/\rmd w_a$ (dashed line), so that a small 
freedom in $\Omega_{\rm K}$ at the percent level can greatly 
degrade the constraint on $w_a$. Similarly, the relatively tighter
constraint on $w_0$ compared to that on $w_a$ with all techniques
can be explained by the relative amplitude of the derivatives.
Since the distance derivative 
$\rmd \ln D_{\rm A}/\rmd \Omega_{\rm K}$
decreases more slowly with increasing redshift than 
$\rmd \ln D_{\rm A}/\rmd w_0$ (solid line)
and $\rmd \ln D_{\rm A}/\rmd w_a$, the degeneracy between dark energy
and curvature can be lifted by extending the measurements to higher 
redshift. This explains the 
factor of $\sim 3$ difference in BAO $\sigma(\Omega_{\rm K})$ 
between the LSST survey and the shallower PS4 survey.
The magnitude of the growth derivative with respect to 
$\Omega_{\rm K}$ (open circles) is at least 20 times larger 
than those with respect to $w_0$ (open squares) and $w_a$ (open
triangles); the relative sign between the growth and distance 
derivatives with respect to $\Omega_{\rm K}$ is opposite to those
with respect to $w_0$ and $w_a$. Consequently, WL, being able to
measure the growth function, can determine the curvature well even
without high-redshift data, so that
the PS4 WL constraint on $\Omega_{\rm K}$ is only moderately 
worse than that of LSST.

Finally, we present the joint constraints of LSST BAO and WL for some 
interesting dark energy models in \fref{fig:eos}. All the models 
assume the same fiducial values for the other 9 cosmological 
parameters, i.e., essentially, the same distance to the last 
scattering surface. This requires modifications to the two implicit
parameters: $\Omega_{\rm X} = 0.61$ and $h = 0.57$ for the 
supergravity-inspired model at $(w_0,w_a) = (-0.8,0.6)$. For the 
models along the dashed line, $\Omega_{\rm X} = 0.76$ and $h = 0.73$
are preserved as well. In other words, $w_0$ and $w_a$ are degenerate 
along the dashed line, if the distance to the last scattering 
surface is 
the only constraint on dark energy. \Fref{fig:eos} demonstrates 
that a joint analysis of LSST BAO and WL can indeed distinguish 
between some dark energy models that are quite degenerate to CMB 
or WL alone. Therefore, it is promising that, with multiple 
techniques as well as various statistics of the same galaxy 
and shear data, we will gain valuable insight to the mystery of 
dark energy from ambitious multiband imaging surveys.

\ack
We thank Gary Bernstein for many helpful comments, and 
Sarah Bridle, Lloyd Knox, 
Michael Schneider, Yong-Seon Song, Anthony Tyson, and David 
Wittman for useful conversations and suggestions. This work
was supported by NSF under Grant No. 
0307961 and 0441072 and NASA under grant No. NAG5-11098.

\appendix

\section{Linear growth function} \label{app:ling}
Assuming that dark energy is spatially homogeneous, we can 
solve the the linear growth function $G(a)$ from the linear 
perturbation equation 
\be \label{eq:growth}
G''(a) + \frac{G'(a)}{2a}\left\{4-\Omega_{\rm m}(a)-
\Omega_{\rm X}(a)\left[1 + 3 w(a)\right]\right\} = 
\frac{3 \Omega_{\rm m}(a)}{2a^2} G(a),
\ee
where $\Omega_{\rm X}(a)$ is the ratio of dark energy density 
to the critical density, and  
\bea \nonumber
\Omega_{\rm m}(a) &=& a^{-3} E^{-2}(a)\, \Omega_{\rm m},
\\ \nonumber
\Omega_{\rm X}(a) &=& E^{-2}(a) \, \Omega_{\rm X} \,
\exp\left\{3\int_a^1\rmd \ln a'\,\left[1 + w(a')\right]\right\},
\\ \nonumber
E^2(a) &=& H_0^{-2} H^2(a) = a^{-3} \Omega_{\rm m} +
a^{-2}(1 - \Omega_{\rm m} - \Omega_{\rm X}) + \\ \nonumber
&& \qquad\qquad\quad \Omega_{\rm X} \, 
\exp\left\{3\int_a^1\rmd \ln a'\,\left[1 + w(a')\right]\right\},
\eea
with $\Omega_{\rm m} \equiv \Omega_{\rm m}(1)$ and 
$\Omega_{\rm X} \equiv \Omega_{\rm X}(1)$. For a flat universe,
\eref{eq:growth} reduces to the from in \cite{wang98}.
With a matter-dominated universe at $a \ll 1$, we set the 
initial condition to $G(a_\rmi) = a_\rmi$, where the 
starting expansion factor satisfies $|G''(a_\rmi)| \ll 1$. 
The radiation component is dropped in the initial condition to 
follow the convention $G(a) = a$ in an Einstein--de Sitter 
universe. This is appropriate since we are only concerned with the
relative growth in matter era.
The linear growth function may be normalised so that $G(1) = 1$ 
regardless the cosmological model in some applications.

\section{Nonlinear matter power spectrum} \label{app:nlps}

Fitting formulae that are calibrated with $N$-body simulations 
\cite{peacock96,smith03} are widely used to map the linear 
matter power spectrum to nonlinear matter power spectrum. We
use the PD96 formula \cite{peacock96} for 
convenience. However, modifications are necessary for the 
following reasons: (1) PD96 formula expects the logarithmic slope 
of the linear power spectrum to be monotonic, which is 
 not satisfied with BAOs, 
and (2) it maps the linear wavenumber to nonlinear 
wavenumber, which would shift the scales of BAOs dramatically. 
To overcome these problems, we use a smooth fitting formula
\cite{eisenstein99a} of the linear power spectrum without BAOs, 
$\Delta_{\rm NB,L}^2(k)$, that otherwise matches the power 
spectrum with BAOs, $\Delta_{\rm BAO,L}^2(k)$, to calculate 
the nonlinear matter power spectrum, $\Delta_{\rm NB}^2(k)$. The 
final nonlinear matter power spectrum with BAOs is then 
$\Delta_{\rm BAO}^2(k) = \Delta_{\rm BAO,L}^2(k) 
\Delta_{\rm NB}^2(k) / \Delta_{\rm NB,L}^2(k)$. We also note 
that one may use third-order perturbation theory to calculate
the nonlinear matter power spectrum with BAOs at $z > 1$
\cite{jeong06}, but there is no clear demonstration that the large 
error of the perturbative calculations at lower redshift 
\cite{jain94} has been rectified.

%\bibliographystyle{iopart-num}
%\bibliography{ref}

\begin{thebibliography}{100}
\expandafter\ifx\csname url\endcsname\relax
  \def\url#1{{\tt #1}}\fi
\expandafter\ifx\csname urlprefix\endcsname\relax\def\urlprefix{URL }\fi
\providecommand{\eprint}[2][]{\url{#2}}
% Bibliography created with iopart-num.bst, v1.0
% hacked by Hu Zhan 8/4/06

\bibitem{riess98}
{Riess} A~G {\em et~al.\/},  1998 {\em \aj\/} {\bf 116} 1009--1038

\bibitem{perlmutter99a}
{Perlmutter} S {\em et~al.\/},  1999 {\em \apj\/} {\bf 517} 565--586

\bibitem{tonry03}
{Tonry} J~L {\em et~al.\/},  2003 {\em \apj\/} {\bf 594} 1--24

\bibitem{riess04}
{Riess} A~G {\em et~al.\/},  2004 {\em \apj\/} {\bf 607} 665--687

\bibitem{dodelson00}
{Dodelson} S and {Knox} L,  2000 {\em Physical Review Letters\/} {\bf 84}
  3523--3526

\bibitem{percival02}
{Percival} W~J {\em et~al.\/},  2002 {\em \mnras\/} {\bf 337} 1068--1080

\bibitem{spergel03}
{Spergel} D~N {\em et~al.\/},  2003 {\em \apjs\/} {\bf 148} 175--194

\bibitem{tegmark04b}
{Tegmark} M {\em et~al.\/},  2004 {\em \prd\/} {\bf 69}(10) 103501--+

\bibitem{spergel06}
{Spergel} D~N {\em et~al.\/},  2006 \textit{Preprint} \eprint{astro-ph/0603449}

\bibitem{peebles70}
{Peebles} P~J~E and {Yu} J~T,  1970 {\em \apj\/} {\bf 162} 815--+

\bibitem{bond84}
{Bond} J~R and {Efstathiou} G,  1984 {\em \apjl\/} {\bf 285} L45--L48

\bibitem{holtzman89}
{Holtzman} J~A,  1989 {\em \apjs\/} {\bf 71} 1--24

\bibitem{hu99a}
{Hu} W and {Tegmark} M,  1999 {\em \apjl\/} {\bf 514} L65--L68

\bibitem{hu02b}
{Hu} W,  2002 {\em \prd\/} {\bf 66}(8) 083515--+

\bibitem{huterer02}
{Huterer} D,  2002 {\em \prd\/} {\bf 65}(6) 063001--+

\bibitem{refregier03}
{Refregier} A,  2003 {\em \araa\/} {\bf 41} 645--668

\bibitem{ishak04a}
{Ishak} M, {Hirata} C~M, {McDonald} P and {Seljak} U,  2004 {\em \prd\/} {\bf
  69}(8) 083514--+

\bibitem{takada04b}
{Takada} M and {Jain} B,  2004 {\em \mnras\/} {\bf 348} 897--915

\bibitem{song04}
{Song} Y~S and {Knox} L,  2004 {\em \prd\/} {\bf 70}(6) 063510--+

\bibitem{knox06b}
{Knox} L, {Song} Y~S and {Tyson} J~A,  2006 {\em \prd\/} {\bf 74}(2) 023512--+

\bibitem{knox06c}
{Knox} L, {Song} Y~S and {Zhan} H,  2006  
\textit{Preprint}  \eprint{astro-ph/0605536}

\bibitem{haiman01}
{Haiman} Z, {Mohr} J~J and {Holder} G~P,  2001 {\em \apj\/} {\bf 553} 545--561

\bibitem{hu03a}
{Hu} W,  2003 {\em \prd\/} {\bf 67}(8) 081304--+

\bibitem{lima04}
{Lima} M and {Hu} W,  2004 {\em \prd\/} {\bf 70}(4) 043504--+

\bibitem{lima05}
{Lima} M and {Hu} W,  2005 {\em \prd\/} {\bf 72}(4) 043006--+

\bibitem{eisenstein98}
{Eisenstein} D~J, {Hu} W and {Tegmark} M,  1998 {\em \apjl\/} {\bf 504} L57

\bibitem{cooray01b}
{Cooray} A, {Hu} W, {Huterer} D and {Joffre} M,  2001 {\em \apjl\/} {\bf 557}
  L7--L10

\bibitem{bartelmann01}
{Bartelmann} M and {Schneider} P,  2001 {\em \physrep\/} {\bf 340} 291--472

\bibitem{blake03}
{Blake} C and {Glazebrook} K,  2003 {\em \apj\/} {\bf 594} 665--673

\bibitem{hu03b}
{Hu} W and {Haiman} Z,  2003 {\em \prd\/} {\bf 68}(6) 063004--+

\bibitem{linder03}
{Linder} E~V,  2003 {\em \prd\/} {\bf 68}(8) 083504--+

\bibitem{seo03}
{Seo} H and {Eisenstein} D~J,  2003 {\em \apj\/} {\bf 598} 720--740

\bibitem{hu04b}
{Hu} W and {Jain} B,  2004 {\em \prd\/} {\bf 70}(4) 043009--+

\bibitem{angulo05}
{Angulo} R, {Baugh} C~M, {Frenk} C~S, {Bower} R~G, {Jenkins} A and {Morris}
  S~L,  2005 {\em \mnras\/} {\bf 362} L25--L29

\bibitem{blake05}
{Blake} C and {Bridle} S,  2005 {\em \mnras\/} {\bf 363} 1329--1348

\bibitem{glazebrook05}
{Glazebrook} K and {Blake} C,  2005 {\em \apj\/} {\bf 631} 1--20

\bibitem{seo05}
{Seo} H~J and {Eisenstein} D~J,  2005 {\em \apj\/} {\bf 633} 575--588

\bibitem{white05}
{White} M,  2005 {\em Astroparticle Physics\/} {\bf 24} 334--344

\bibitem{bernstein06}
{Bernstein} G,  2006 {\em \apj\/} {\bf 637} 598--607

\bibitem{dolney06}
{Dolney} D, {Jain} B and {Takada} M,  2006 {\em \mnras\/} {\bf 366} 884--898

\bibitem{eisenstein06a}
{Eisenstein} D~J, {Seo} H~J and {White} M,  2006 \textit{Preprint}
  \eprint{astro-ph/0604361}

\bibitem{eisenstein06b}
{Eisenstein} D~J, {Seo} H~J, {Sirko} E and {Spergel} D,  2006 \textit{Preprint}
  \eprint{astro-ph/0604362}

\bibitem{wang06}
{Wang} Y,  2006 {\em \apj\/} {\bf 647} 1--7

\bibitem{zhan06c}
{Zhan} H and {Knox} L,  2006 {\em \apj\/} {\bf 644} 663--670

\bibitem{chevallier01}
{Chevallier} M and {Polarski} D,  2001 {\em International Journal of Modern
  Physics D\/} {\bf 10} 213--223

\bibitem{bernstein04}
{Bernstein} G and {Jain} B,  2004 {\em \apj\/} {\bf 600} 17--25

\bibitem{ishak05}
{Ishak} M and {Hirata} C~M,  2005 {\em \prd\/} {\bf 71}(2) 023002--+

\bibitem{ma06}
{Ma} Z, {Hu} W and {Huterer} D,  2006 {\em \apj\/} {\bf 636} 21--29

\bibitem{huterer06}
{Huterer} D, {Takada} M, {Bernstein} G and {Jain} B,  2006 {\em \mnras\/} {\bf
  366} 101--114

\bibitem{white04}
{White} M,  2004 {\em Astroparticle Physics\/} {\bf 22} 211--217

\bibitem{zhan04c}
{Zhan} H and {Knox} L,  2004 {\em \apjl\/} {\bf 616} L75--L78

\bibitem{hagan05}
{Hagan} B, {Ma} C~P and {Kravtsov} A~V,  2005 {\em \apj\/} {\bf 633} 537--541

\bibitem{heitmann05}
{Heitmann} K, {Ricker} P~M, {Warren} M~S and {Habib} S,  2005 {\em \apjs\/}
  {\bf 160} 28--58

\bibitem{zhan06a}
{Zhan} H,  2006 {\em \apj\/} {\bf 639} 617--620

\bibitem{huterer05a}
{Huterer} D and {Takada} M,  2005 {\em Astroparticle Physics\/} {\bf 23}
  369--376

\bibitem{ratra88}
{Ratra} B and {Peebles} P~J~E,  1988 {\em \prd\/} {\bf 37} 3406--3427

\bibitem{huey99}
{Huey} G, {Wang} L, {Dave} R, {Caldwell} R~R and {Steinhardt} P~J,  1999 {\em
  \prd\/} {\bf 59}(6) 063005--+

\bibitem{perlmutter99b}
{Perlmutter} S, {Turner} M~S and {White} M,  1999 {\em Physical Review
  Letters\/} {\bf 83} 670--673

\bibitem{wang00}
{Wang} L, {Caldwell} R~R, {Ostriker} J~P and {Steinhardt} P~J,  2000 {\em
  \apj\/} {\bf 530} 17--35

\bibitem{dvali00}
{Dvali} G, {Gabadadze} G and {Porrati} M,  2000 {\em Physics Letters B\/} {\bf
  485} 208--214

\bibitem{deffayet01}
{Deffayet} C,  2001 {\em Physics Letters B\/} {\bf 502} 199--208

\bibitem{abazajian03}
{Abazajian} K and {Dodelson} S,  2003 {\em Physical Review Letters\/} {\bf
  91}(4) 041301--+

\bibitem{linder04}
{Linder} E~V,  2004 {\em \prd\/} {\bf 70}(2) 023511--+

\bibitem{simpson05}
{Simpson} F and {Bridle} S,  2005 {\em \prd\/} {\bf 71}(8) 083501--+

\bibitem{song05}
{Song} Y~S,  2005 {\em \prd\/} {\bf 71}(2) 024026--+

\bibitem{zhang05}
{Zhang} J, {Hui} L and {Stebbins} A,  2005 {\em \apj\/} {\bf 635} 806--820

\bibitem{bacon00}
{Bacon} D~J, {Refregier} A~R and {Ellis} R~S,  2000 {\em \mnras\/} {\bf 318}
  625--640

\bibitem{wittman00}
{Wittman} D~M, {Tyson} J~A, {Kirkman} D, {Dell'Antonio} I and {Bernstein} G,
  2000 {\em \nat\/} {\bf 405} 143--148

\bibitem{vanwaerbeke01}
{Van Waerbeke} L {\em et~al.\/},  2001 {\em \aap\/} {\bf 374} 757--769

\bibitem{jarvis03}
{Jarvis} M, {Bernstein} G~M, {Fischer} P, {Smith} D, {Jain} B, {Tyson} J~A and
  {Wittman} D,  2003 {\em \aj\/} {\bf 125} 1014--1032

\bibitem{hoekstra05}
{Hoekstra} H, {Mellier} Y, {van Waerbeke} L, {Semboloni} E, {Fu} L, {Hudson}
  M~J, {Parker} L~C, {Tereno} I and {Benabed} K,  2006 {\em \apj\/} {\bf 647}
  116--127

\bibitem{eisenstein05}
{Eisenstein} D~J {\em et~al.\/},  2005 {\em \apj\/} {\bf 633} 560--574

\bibitem{cole05}
{Cole} S {\em et~al.\/},  2005 {\em \mnras\/} {\bf 362} 505--534

\bibitem{huetsi06}
{H{\"u}tsi} G,  2006 {\em \aap\/} {\bf 449} 891--902

\bibitem{ilbert06}
{Ilbert} O {\em et~al.\/},  2006 \textit{Preprint} \eprint{astro-ph/0603217}

\bibitem{padmanabhan05}
{Padmanabhan} N {\em et~al.\/},  2005 {\em \mnras\/} {\bf 359} 237--250

\bibitem{padmanabhan06}
{Padmanabhan} N {\em et~al.\/},  2006 \textit{Preprint}
  \eprint{astro-ph/0605302}

\bibitem{blake06b}
{Blake} C, {Collister} A, {Bridle} S and {Lahav} O,  2006 \textit{Preprint}
  \eprint{astro-ph/0605303}

\bibitem{schneider98}
{Schneider} P,  1998 {\em \apj\/} {\bf 498} 43--+

\bibitem{yee98}
{Yee} H~K~C,  1998 \textit{Preprint} \eprint{astro-ph/9809347}

\bibitem{connolly97}
{Connolly} A~J, {Szalay} A~S, {Dickinson} M, {Subbarao} M~U and {Brunner} R~J,
  1997 {\em \apjl\/} {\bf 486} L11+

\bibitem{benitez00}
{Ben{\'{\i}}tez} N,  2000 {\em \apj\/} {\bf 536} 571--583

\bibitem{fontana00}
{Fontana} A, {D'Odorico} S, {Poli} F, {Giallongo} E, {Arnouts} S, {Cristiani}
  S, {Moorwood} A and {Saracco} P,  2000 {\em \aj\/} {\bf 120} 2206--2219

\bibitem{babbedge04}
{Babbedge} T~S~R, {Rowan-Robinson} M, {Gonzalez-Solares} E, {Polletta} M,
  {Berta} S, {P{\'e}rez-Fournon} I, {Oliver} S, {Salaman} D~M, {Irwin} M and
  {Weatherley} S~J,  2004 {\em \mnras\/} {\bf 353} 654--672

\bibitem{mobasher04}
{Mobasher} B {\em et~al.\/},  2004 {\em \apjl\/} {\bf 600} L167--L170

\bibitem{schneider06}
{Schneider} M, {Knox} L, {Zhan} H and {Connolly} A,  2006 \textit{Preprint}
  \eprint{astro-ph/0606098}

\bibitem{newman06}
{Newman} J,  2006 In preparation

\bibitem{limber54}
{Limber} D~N,  1954 {\em \apj\/} {\bf 119} 655--+

\bibitem{kaiser92}
{Kaiser} N,  1992 {\em \apj\/} {\bf 388} 272--286

\bibitem{verde02}
{Verde} L {\em et~al.\/},  2002 {\em \mnras\/} {\bf 335} 432--440

\bibitem{tegmark04a}
{Tegmark} M {\em et~al.\/},  2004 {\em \apj\/} {\bf 606} 702--740

\bibitem{tegmark02}
{Tegmark} M {\em et~al.\/},  2002 {\em \apj\/} {\bf 571} 191--205

\bibitem{guzik05}
{Guzik} J and {Bernstein} G,  2005 {\em \prd\/} {\bf 72}(4) 043503--+

\bibitem{jain06}
{Jain} B, {Jarvis} M and {Bernstein} G,  2006 {\em Journal of Cosmology and
  Astro-Particle Physics\/} JCAP02(2006)001

\bibitem{zaldarriaga00}
{Zaldarriaga} M and {Seljak} U,  2000 {\em \apjs\/} {\bf 129} 431--434

\bibitem{peacock96}
{Peacock} J~A and {Dodds} S~J,  1996 {\em \mnras\/} {\bf 280} L19--L26

\bibitem{zaldarriaga97}
{Zaldarriaga} M, {Spergel} D~N and {Seljak} U,  1997 {\em \apj\/} {\bf 488}
  1--+

\bibitem{eisenstein99b}
{Eisenstein} D~J, {Hu} W and {Tegmark} M,  1999 {\em \apj\/} {\bf 518} 2--23

\bibitem{kaplinghat03}
{Kaplinghat} M, {Knox} L and {Song} Y~S,  2003 {\em Physical Review Letters\/}
  {\bf 91}(24) 241301--+

\bibitem{sacks67}
{Sachs} R~K and {Wolfe} A~M,  1967 {\em \apj\/} {\bf 147} 73--+

\bibitem{jungman96}
{Jungman} G, {Kamionkowski} M, {Kosowsky} A and {Spergel} D~N,  1996 {\em
  \prd\/} {\bf 54} 1332--1344

\bibitem{vogeley96}
{Vogeley} M~S and {Szalay} A~S,  1996 {\em \apj\/} {\bf 465} 34--+

\bibitem{tegmark97a}
{Tegmark} M, {Taylor} A~N and {Heavens} A~F,  1997 {\em \apj\/} {\bf 480} 22--+

\bibitem{tegmark97b}
{Tegmark} M,  1997 {\em Physical Review Letters\/} {\bf 79} 3806--3809

\bibitem{cooray01a}
{Cooray} A and {Hu} W,  2001 {\em \apj\/} {\bf 554} 56--66

\bibitem{weinberg04}
{Weinberg} D~H, {Dav{\' e}} R, {Katz} N and {Hernquist} L,  2004 {\em \apj\/}
  {\bf 601} 1--21

\bibitem{hoekstra02}
{Hoekstra} H, {van Waerbeke} L, {Gladders} M~D, {Mellier} Y and {Yee} H~K~C,
  2002 {\em \apj\/} {\bf 577} 604--614

\bibitem{seljak05}
{Seljak} U, {Makarov} A, {Mandelbaum} R, {Hirata} C~M, {Padmanabhan} N,
  {McDonald} P, {Blanton} M~R, {Tegmark} M, {Bahcall} N~A and {Brinkmann} J,
  2005 {\em \prd\/} {\bf 71}(4) 043511--+

\bibitem{martin06}
{Martin} D and {Albrecht} A,  2006 \textit{Preprint} \eprint{astro-ph/0604401}

\bibitem{zhan06b}
{Zhan} H, {Knox} L, {Tyson} J~A and {Margoniner} V,  2006 {\em \apj\/} {\bf
  640} 8--17

\bibitem{knox06a}
{Knox} L,  2006 {\em \prd\/} {\bf 73}(2) 023503--+

\bibitem{hu02a}
{Hu} W,  2002 {\em \prd\/} {\bf 65}(2) 023003--+

\bibitem{linder05b}
{Linder} E~V,  2005 {\em Astroparticle Physics\/} {\bf 24} 391--399

\bibitem{wang98}
{Wang} L and {Steinhardt} P~J,  1998 {\em \apj\/} {\bf 508} 483--490

\bibitem{smith03}
{Smith} R~E, {Peacock} J~A, {Jenkins} A, {White} S~D~M, {Frenk} C~S, {Pearce}
  F~R, {Thomas} P~A, {Efstathiou} G and {Couchman} H~M~P,  2003 {\em \mnras\/}
  {\bf 341} 1311--1332

\bibitem{eisenstein99a}
{Eisenstein} D~J and {Hu} W,  1999 {\em \apj\/} {\bf 511} 5--15

\bibitem{jeong06}
{Jeong} D and {Komatsu} E,  2006 \textit{Preprint} \eprint{astro-ph/0604075}

\bibitem{jain94}
{Jain} B and {Bertschinger} E,  1994 {\em \apj\/} {\bf 431} 495--505

\end{thebibliography}
%\end{document}

\providecommand{\newblock}{}

\end{document}